\documentclass[10pt, conference, letterpaper]{IEEEtran}
\usepackage{epsfig,amsmath,hhline}

\usepackage{times}

\usepackage{latexsym}
\usepackage{amsmath}
\usepackage{nccbbb}
\usepackage{amsthm}
\usepackage{url}
\usepackage{algorithm}
\usepackage{algorithmic}
\usepackage{dsfont}
\usepackage{color}
\usepackage{hyperref}

\usepackage{tabularx}



\newtheorem{theorem}{Theorem}
\newtheorem{lemma}{Lemma}

\theoremstyle{definition}

\theoremstyle{plain}
\newtheorem{assumption}{Assumption}

\newtheorem{definition}{Definition}

\theoremstyle{remark}

\newcommand*{\affaddr}[1]{#1} 
\newcommand*{\affmark}[1][*]{\textsuperscript{#1}}

\usepackage{graphicx}
\usepackage{amssymb}
\usepackage{epstopdf}
\title{Dynamic Cloud Network Control under Reconfiguration Delay and Cost}

\author{
Chang-Heng Wang\affmark[1], Jaime Llorca\affmark[2], Antonia M. Tulino\affmark[2], and Tara Javidi\affmark[1]\\
\affaddr{\affmark[1]University of California, San Diego, CA. Email:\{chw009, tjavidi\}@ucsd.edu}\\
\affaddr{\affmark[2]Nokia Bell Labs, NJ. Email:\{jaime.llorca, a.tulino\}@nokia-bell-labs.com}\\
}

\begin{document}

\maketitle

\begin{abstract}
Network virtualization and programmability allow operators to deploy a wide range of services over a common physical infrastructure and elastically allocate cloud and network resources according to changing requirements. While the elastic reconfiguration of virtual resources enables dynamically scaling capacity in order to support service demands with minimal operational cost, reconfiguration operations make resources unavailable during a given time period and may incur additional cost. In this paper, we address the dynamic cloud network control problem under non-negligible reconfiguration delay and cost. We show that while the capacity region remains unchanged regardless of the reconfiguration delay/cost values, a reconfiguration-agnostic policy may fail to guarantee throughput-optimality and minimum cost under nonzero reconfiguration delay/cost. We then present an adaptive dynamic cloud network control policy that allows network nodes to make local flow scheduling and resource allocation decisions while controlling the frequency of reconfiguration in order to support any input rate in the capacity region and achieve arbitrarily close to minimum cost for any finite reconfiguration delay/cost values. 
\end{abstract}

\section{Introduction}

The emergence of network function virtualization (NFV) and software defined networking (SDN) enables network services to be deployed in the form of interconnected software functions instantiated over commercial off-the-shelf servers at multiple cloud locations and interconnected via a programmable network fabric. This allows cloud network operators to host a large variety of services over a common general purpose infrastructure and dynamically allocate resources according to changing demands, reducing both capital and operational expenses.

The unprecendented flexibility of the cloud networking paradigm provides exciting opportunities for future service scenarios and stimulates research in key technical areas such as optimal function placement, service flow routing, and joint cloud/network resource allocation. 
One line of research addressed the virtual network functions placement problem from a static global optimization point of view, in which the goal is to find the placement of virtual functions and the routing of network flows that meet service demands with minimum cost \cite{vnf,csdp,nsdp}. However, requirements for the prior knowledge of global system information and service demands restrict the use of such centralized policies to relatively small-scale scenarios with relatively static demands.
In contrast, recent works have leveraged ideas from dynamic network control to design distributed control policies for computing networks, in which nodes make local decisions on processing and transmission flow scheduling \cite{Destounis_info16}, as well as associated compute and network resource allocation \cite{dcnc_info16,dcnc_icc16}, with global system guarantees. The work in \cite{Destounis_info16} proposes a backpressure-based algorithm for maximizing the rate of queries for a computation operation on remote data, while \cite{dcnc_info16,dcnc_icc16} present cloud network control policies for service function chains that guarantee throughput-optimality and minimun average cloud network cost. 
While the dynamic cloud network control (DCNC) algorithm presented in \cite{dcnc_icc16} shows promise in serving varying workloads with minimum cost by dynamically adjusting resource allocation and scheduling decisions, it overlooks the fact that the reconfiguration of virtual compute and network resources takes a non-negligible amount of time and may incur additional cost. As an example, starting up a virtual machine (VM) can take up to 5-10 minutes~\cite{vm_2012}. 
A control policy that is unaware of the reconfiguration delay and cost associated with the cloud and network resources, may perform excessive reconfigurations that can lead to increased congestion and overall operational cost. 

The reconfiguration delay associated with flow scheduling has been studied in the context of the switch model~\cite{SCB, MWMH, AMW_ToN}, and multi-hop networks~\cite{Adaptive_Backpressure, BMP}, and signal control in transportation systems~\cite{BMP}. In these works, throughput optimal scheduling policies under any finite reconfiguration delay have been proposed. However, resource allocation, and thus cost minimization, is not considered in the settings of these works. Regarding reconfiguration cost,~\cite{reconfiguration_sdn} addressed the cost of flow reconfigurations in SDN by designing a control policy that minimizes total flow allocation cost subject to a given reconfiguration cost budget. In~\cite{maxweight_switching_cost}, the reconfiguration cost associated with switching base stations on and off in a dynamic wireless network setting was considered. The proposed approach requires arrival and channel statistics for activation decisions, and leverages an explore-exploit policy in the case that this information is not available. 

In this paper, we address the problem of optimal control of multi-hop multi-commodity cloud networks in practical settings characterized by non-negligible reconfiguration delay and cost. 
The contributions of this work can be summarized as follows: 

\begin{itemize}
\item[(i)] We show that the capacity region and the minimum time average cost remains the same even in the presence of reconfiguration delay and cost, given that reconfiguration delay and cost values are finite. 

\item[(ii)] We show that a reconfiguration-agnostic policy that is throughput optimal and achieves arbitrarily close to the minimum time average cost in the no reconfiguration delay/cost regime does not necessarily retain these properties when reconfiguration delay/cost exists. 

\item[(iii)] We propose a distributed flow scheduling and resource allocation policy that is able to guarantee cloud network throughput and cost optimality for any finite values of reconfiguration delay/cost. The proposed Adaptive Dynamic Cloud Network Control (ADCNC) policy adapts the frequency of reconfiguration utilizing the queue length information, and does not require prior knowledge on arrival statistics or the exact value of the reconfiguration overheads. 

\item[(iv)] The problem considered in this work is a combination of the cost-minimizing flow scheduling and the multi-hop scheduling with reconfiguration delay, where known solutions of each does not trivially apply in this generalization. The proposed ADCNC policy generalizes the applicability of adaptive policy to incorporate the regime of cost-minimizing flow scheduling, which requires an appropriate modification in the reconfiguration criterion.

\end{itemize}

The rest of the paper is organized as follows. We introduce the system model and formulate the cloud network control problem in Section~\ref{Sec:model}. 
With the formulated problem, we compare our setting to the existing literature in Section~\ref{Sec:related}.
The impact of the reconfiguration delay/cost is illustrated in Section~\ref{Sec:impact} to motivate the problem considered in this work.
In Section~\ref{Sec:policy}, we introduce ADCNC policy, and characterize its performance guarantee. Simulation results are presented in Section~\ref{Sec:simulation}. We then conclude with some discussions and future directions in Section~\ref{Sec:conclusion}.

\textsl{Notation:} Throughout the paper, we use $\mathds{1}_{\{\cdot\}}$ to denote the indicator function, and $|\mathcal{S}|$ to denote the cardinality of a set $\mathcal{S}$. We also use $[x]^+$ as a shorthand for $\max\{x, 0\}$.

\section{System Model}\label{Sec:model}

A cloud network is modeled as a directed graph $\mathcal{G} = (\mathcal{V}, \mathcal{E})$, with $|\mathcal{V}| = V$ vertices and $|\mathcal{E}| = E$ edges representing cloud nodes and network links, respectively. 
Cloud and network resources are characterized by their processing and transmission capacities and costs, as follows:
\begin{itemize}
    \item $\mathcal{K}_{i} = \{0, 1, \dots, K_i\}$: the set of possible processing resource units at node $i \in \mathcal{V}$
    \item $\mathcal{K}_{ij} = \{0, 1, \dots, K_{ij}\}$: the set of possible transmission resource units at link $(i,j) \in \mathcal{E}$ 
    \item $C_{i}(k)$: the processing capacity resulting from the allocation of $k$ processing resource units at node $i \in \mathcal{V}$
    \item $C_{ij}(k)$: the transmission capacity resulting from the allocation of $k$ transmission resource units at link $(i,j) \in \mathcal{E}$ 
    \item $w_{i}(k)$: the cost of maintaining $k$ processing resource units at node $i \in \mathcal{V}$
    \item $w_{ij}(k)$: the cost of maintaining $k$ transmission resource units at link $(i,j) \in \mathcal{E}$ 
    \item $e_{i}$: the cost per processing flow unit at node $i \in \mathcal{V}$
    \item $e_{ij}$: the cost per transmission flow unit at link $(i,j) \in \mathcal{E}$ 
\end{itemize}

Throughout the rest of the discussion, we make the following assumption on the capacities and costs of cloud and network resources: 
\begin{assumption}
\label{assumption_link}
For any node $i \in \mathcal{V}$ and any link $(i,j) \in \mathcal{E}$, we assume that both the capacity and the cost are strictly increasing with the amount of resource assigned. In other words, given any node $i$, for any $k$ such that $0 \leq k \leq K_{i}-1$, we have $C_{i}(k) < C_{i}(k+1)$ and $w_{i}(k) < w_{i}(k+1)$; similarly, given any link $(i,j)$, for any $k$ such that $0 \leq k \leq K_{ij}-1$, we have $C_{ij}(k) < C_{ij}(k+1)$ and $w_{ij}(k) < w_{ij}(k+1)$.
\end{assumption}

\subsection{Service model}
Cloud network $\mathcal G$ offers a set of services $\Phi$. Each service $\phi \in \Phi$ is described by a chain of service functions. 
We let $\mathcal{M}_{\phi} = \{1, 2, \dots, M_{\phi}\}$ denote the ordered set of functions of service $\phi$, hence the tuple $(\phi,m)$ represents the $m$-th function of service $\phi$. 

In order to describe the flow of packets through a service chain, we adopt a multi-commodity-chain flow model as in~\cite{csdp,dcnc_info16,dcnc_icc16}, in which a commodity represents the flow of packets at a given stage of a service chain. 
In particular, a commodity-$c$ flow is specified by source node $s_c$, destination node $d_c$, and function $(\phi,m)_c$, indicating the flow of packets with origin at $s_c$ and destination at $d_c$ that have been processed by the first $m$ functions of service $\phi$. 
For ease of exposition, we let $c^+$ and $c^-$ denote the commodities that succeed and preceed commodity $c$ in its service chain, respectively.

Each service function has potentially distinct processing requirement, which may also vary between cloud locations. We let $\rho_{i}^{(c)}$ denote the processing-transmission flow ratio of function $(\phi,m)_c$ at node $i$. That is, when one transmission flow unit of commodity $c$ goes through function $(\phi,m)_c$ at node $i$, it occupies $\rho_{i}^{(c)}$ processing flow units.
In addition, our service model also captures the possibility of flow scaling. We denote by $\xi^{(c)} > 0$ the scaling factor of function $(\phi,m)_c$, indicating that function $(\phi,m)_c$ generates an average of $\xi^{(c)}$ output packets of commodity $c$ per input packet of commodity $c^-$.

\subsection{Reconfiguration Delay and Cost}

We consider cloud network control policies that adjust the configuration of cloud and network resources, as well as the schedule of commodity flows, according to changing demands. We assume that such reconfigurations may incur the following two types of overhead:
\begin{itemize}
\item Reconfiguration delay (time): This is the time duration for the reconfiguration process to complete. We assume that during the reconfiguration process, the associated function (transmission or processing of commodity flows) is not available. We denote by $\delta_{i}$ the reconfiguration delay for node $i \in \mathcal{V}$, and by $\delta_{ij}$ the reconfiguration delay for link $(i,j) \in \mathcal{E}$.

\item Reconfiguration cost: This is the cost/penalty associated with each reconfiguration operation. Let $\eta_{i}$ denote the reconfiguration cost for node $i \in \mathcal{V}$, and $\eta_{ij}$ denote the reconfiguration cost for link $(i,j) \in \mathcal{E}$.
\end{itemize}
In the rest of the paper, we use $\Delta$ to denote the reconfiguration delay and cost structure of a cloud network, where
$\Delta = \Big\{ \{\delta_{i}\}_{i\in\mathcal{V}}, \{\delta_{ij}\}_{(i,j)\in\mathcal{E}}, \{\eta_{i}\}_{i\in\mathcal{V}}, \{\eta_{ij}\}_{(i,j)\in\mathcal{E}} \Big\}$.

We consider a time slotted system with slots normalized to integral units $t \in \{0,1,2,\dots\}$. 
Suppose that node $i$ reconfigures the processing resource allocation or the commodity being processed at time $t$. Then, 
flow processing at node $i$ becomes unavailable during time period $[t, t+\delta_{i}]$, and a reconfiguration cost $\eta_{i}$ is incurred at time $t$. Similarly, suppose that link $(i,j)$ reconfigures the transmission resource allocation or the commodity being transmitted at time $t$. Then, flow tranmission is unavailable during $[t, t+\delta_{ij}]$, and a reconfiguration cost $\eta_{ij}$ is incurred at time $t$.

Note that we consider a worst-case reconfiguration delay model in that we assume complete unavailability of packet processing or transmission functionality at a node or link undergoing reconfiguration. 
Importantly, a throughput-optimal policy for this worst-case reconfiguration delay model will guarantee throughput-optimality for any other less restrictive model. 
Extensions to this model are discussed in Section \ref{Sec:extension}.

For ease of discussion in the following, we also define $r_{i}(t)$ and $r_{ij}(t)$ to denote the reconfiguration status:
\begin{itemize}
\item $r_{i}(t)$: the time remaining in the reconfiguration process at node $i \in \mathcal{V}$
\item $r_{ij}(t)$: the time remaining in the reconfiguration process at link $(i,j) \in \mathcal{E}$, where $i,j \in \mathcal{V}$
\end{itemize}
By definition, these processes evolve as follows: At any time $t$, if node $i$ (or link $(i,j)$) reconfigures, then set $r_i(t) = \delta_{i}$ (or $r_{ij}(t) = \delta_{ij}$, respectively); otherwise, set $r_i(t) = [r_i(t-1) - 1]^+$ (or $r_{ij}(t) = [r_{ij}(t-1) - 1]^+$, respectively).

\subsection{Queueing Model}
Let $Q_{i}^{(c)}(t)$ denote the queue backlog of commodity-$c$ packets at node $i$ at the beginning of time slot $t$. We denote by $a_{i}^{(c)}(t)$ the exogenous arrivals of commodity-$c$ packets at node $i$ during time slot $t$. Throughout this paper, we make the following assumptions for the exogenous arrival processes. 

\begin{assumption}
Each exogenous arrival process is independent and identically distributed (i.i.d.) over time, with $\bbbe[a_{i}^{(c)}(t)] = \lambda_{i}^{(c)}$. Furthermore, each exogenous arrival process has bounded support. In other words, there exist $a_{\max} < \infty$ such that $a_{i}^{(c)}(t) \leq a_{\max}$, $\forall i \in \mathcal{V}, c \in \mathcal{C}$, $\forall t$.
\label{assumption_arrivals}
\end{assumption}
\vskip -0.1in
At each time slot $t$, each node makes the following transmission and processing  scheduling and resource allocation decisions:
\begin{itemize}
\item $\mu_{i}^{(c)}(t)$: the flow rate of commodity $c$ being processed at node $i$ at time $t$
\item $\mu_{ij}^{(c)}(t)$: the flow rate of commodity $c$ on link $(i,j)$ at time $t$
\item $k_{i}(t)$: the number of processing resource units allocated to node $i$ at time $t$
\item $k_{ij}(t)$: the number of transmission resource units allocated to link $(i,j)$ at time $t$
\end{itemize}

With the aforementioned setup, we may write the queue dynamics for each commodity $c \in \mathcal{C}$ at each node $i$:
\vskip -0.2in
\begin{align}
&Q_{i}^{(c)}(t+1) 
= \bigg[ Q_{i}^{(c)}(t) - \sum_{j \in \mathcal V^{\!+\!}(i)} \mu_{ij}^{(c)}(t) \mathds{1}_{\{r_{ij}(t) = 0\}}  \nonumber \\
&\ \ \ \ \ \ \ \ \ \ \ \ \ \ \ \ \ \ - \mu_{i}^{(c)}(t)  \mathds{1}_{\{r_{i}(t) = 0\}} \bigg]^+  \nonumber \\
&+ \sum_{j \in \mathcal V^{\!-\!}(i)} \mu_{ji}^{(c)}(t) \mathds{1}_{\{r_{ji}(t) = 0\}} + \xi^{(c)} \mu_{i}^{(c^-)}(t)  \mathds{1}_{\{r_{i}(t) = 0\}} + a_{i}^{(c)}(t), 
\label{queue_dynamics}
\end{align}
where $\mathcal V^+(i)$ and $\mathcal V^-(i)$ denote the set of outgoing and incoming neighbors of node $i$, respectively. 

Observe from \eqref{queue_dynamics} that the serving rate of the queue of commodity $c$ at node $i$ is composed of the transmission rate of commodity $c$ of all outgoing links and the local processing rate of commodity $c$. On the other hand, the arrival rate 
is composed of the transmission rate of commodity $c$ of all incoming links and the local processing rate of the preceeding commodity in the service chain $c^-$. 
It is important to note that there is no contribution to both serving and arrival rates from those transmission and processing resources that are undergoing reconfiguration (i.e., $r_{ij}(t)>0$ or $r_{i}(t)>0$), indicating the inability to transmit or process packets during the reconfiguration process.

\subsection{Problem Formulation}

Given a set of service demands with average input rate matrix $\boldsymbol{\lambda} = \{\lambda_{i}^{(c)}\}_{i\in\mathcal{V}, c\in\mathcal{C}}$, the goal 
is to support the demand while minimizing the average cloud network cost. 

In order to formalize the problem, we first introduce the following notion of rate stability, which dictates the ability of a cloud network control policy to support the demand:
\vskip -0.2in
\begin{definition}
A cloud network is \textbf{rate stable} if
\begin{align}
\lim_{t \rightarrow \infty} \frac{Q_{i}^{(c)}}{t} = 0 \ \ \mbox{with prob.} 1, \ \ \forall i\in\mathcal{V}, c\in\mathcal{C}
\end{align}
\end{definition}

With the notion of rate stability, we may then define the capacity of a cloud network and the throughput optimality of a cloud network control policy as follows:

\begin{definition}
For a given cloud network, the \textbf{capacity region} of the cloud network is defined as the closure of all input rate matrix $\boldsymbol{\lambda} = \{\lambda_{i}^{(c)}\}_{i\in\mathcal{V}, c\in\mathcal{C}}$ that could be rate stable under some cloud network control policy.
\end{definition}

\begin{definition}
A cloud network control policy for a cloud network is \textbf{throughput optimal} if the cloud network operated under the control policy is rate stable for any input rate matrix in the capacity region.
\end{definition}

Besides the ability to support the demand, the total operational cost of a cloud network is of concern in many practical settings.
The total cloud network cost consists of the total processing and transmission cost. 
We assume that when a processing/transmission resource is undergoing reconfiguration, processing/transmission allocation cost is not incurred since the resource is not operative until the reconfiguration process completes. 
Hence, we can write the total cloud network cost at time $t$ as
\begin{align}
h(t) 
&= \sum_{i \in \mathcal{V}} \Big( (e_i \mu_i(t) + w_{i}(k_i(t)) ) \mathds{1}_{\{r_{i}(t) = 0\}}   \nonumber \\
& \ \ \ \ \ \ \ \ \ \ \  + \eta_{i} \mathds{1}_{\{(\boldsymbol{\mu}_{i}(t), k_{i}(t)) \neq (\boldsymbol{\mu}_{i}(t-1), k_{i}(t-1))\}} \Big) \nonumber \\
&+ \sum_{(i,j) \in \mathcal{E}} \Big( (e_{ij} \mu_{ij}(t) + w_{ij}(k_{ij}(t)) ) \mathds{1}_{\{r_{ij}(t) = 0\}}  \nonumber \\
& \ \ \ \ \ \ \ \ \ \ \  + \eta_{ij} \mathds{1}_{\{(\boldsymbol{\mu}_{ij}(t), k_{ij}(t)) \neq (\boldsymbol{\mu}_{ij}(t-1), k_{ij}(t-1))\}} \Big) 
\label{network_cost}
\end{align}

We then formulate the dynamic cloud network control problem under reconfiguration delay/cost as the following. Given an input rate matrix $\boldsymbol{\lambda}$ in the capacity region:
\begin{subequations}
\begin{align}
\min \  &\limsup_{t \rightarrow \infty} \frac{1}{t} \sum_{\tau=0}^{t-1} \bbbe[h(\tau)] 
\label{dsdp_minimization} \\
\mbox{s.t. } \ &\mbox{The cloud network is rate stable with input rate $\boldsymbol{\lambda}$} \nonumber \\ 
&\mbox{and under queue length dynamics~(\ref{queue_dynamics})}  \\
& \sum_{c\in\mathcal{C}} \mu_{i}^{(c)}(\tau) \leq \mu_{i}(\tau) \leq C_{i}(k_{i}(\tau)), \qquad\quad \ \ \ \forall i, \tau \\
& \sum_{c\in\mathcal{C}} \mu_{ij}^{(c)}(\tau)  \leq \mu_{ij}(\tau) \leq C_{ij}(k_{ij}(\tau)), \quad \ \forall (i,j), \tau \\
& \mu_{i}^{(c)}(\tau) \geq 0, \ \mu_{ij}^{(c)}(\tau) \geq 0, \qquad\qquad  \forall i, (i,j), c, \tau \\
& 0 \leq k_{i}(\tau) \leq K_{i}, \ 0 \leq k_{ij}(\tau) \leq K_{ij}, \ \ \forall i, (i,j), \tau 
\label{dsdp_constraint}
\end{align}
\end{subequations}

\section{Related Work}\label{Sec:related}

Given the cloud network control with reconfiguration overhead defined in the previous section, we now discuss the relation between the current work and the related literature. 

In the NFV literature, many works consider the cloud network planning problem in the static (or quasi-static) regime. In this regime, the traffic demands are assumed to be fixed data demands, and the proposed approaches focus on optimization through network function placement and flow allocation. 
A thorough comparison of the static NFV placement and routing could be found in~\cite{static_survey}.
In the static regime, the underlying assumption of slow (or no) variation in traffic demands allows the proposed approaches to ignore the reconfiguration overhead as the reconfiguration typically occurs in longer time scale. However, the very same assumption limits the applicability of these approaches in many practical settings where traffic demands constantly change over time.

On the other hand, Dynamic Cloud Network Control (DCNC) policy~\cite{dcnc_icc16} addresses the dynamic setting of cloud network problem. 
It is shown in~\cite{dcnc_icc16} that DCNC policy is throughput optimal and could achieve minimum mean cloud network cost under zero reconfiguration overhead. However, under the practical setting where reconfiguration overhead exists, DCNC policy may suffer serious performance degradation since it is unaware of the reconfiguration overhead. When reconfiguration delay is not negligible, since DCNC policy has no control on the frequency of reconfiguration, it would waste too much time in the reconfiguration delay and lose throughput optimality.


While DCNC policy ignores the practical reconfiguration overhead issue, we note that there are works in the literature that deal with scheduling / control problems under reconfiguration delay~\cite{AMW_ToN, Adaptive_Backpressure, BMP}. For example, Adaptive MaxWeight policy~\cite{AMW_ToN} is proposed for input-queued switch model with reconfiguration delay; on the other hand, Adaptive Backpressure policy is a distributed policy proposed for multi-hop networks with reconfiguration delay. These adaptive policies utilize queue lengths information to determine appropriate timing for reconfiguration, and thus implicitly adapt the frequency of reconfiguration through queue conditions. 

Note that in all the above works that address the reconfiguration delay, the cost minimization is not taken into account. Therefore, we may view these approaches as solutions to a special case of the cloud network control with reconfiguration overhead problem. One of the main contribution of this work is to extend the applicability of the adaptive policies in order to incorporate the cloud network cost minimization.

\section{Impact of Reconfiguration Delay/Cost} \label{Sec:impact}

In this section, we discuss the impact of reconfiguration delay and cost on the performance of a cloud network control policy that is unaware of such reconfiguration delay/cost.

In order to formalize the notion of performance measure, we start with the characterization of the cloud network capacity region and the minimum average cloud network cost required for network stability. 
The cloud network capacity region $\boldsymbol{\Lambda}_{\Delta}$ is defined as the closure of all input rate matrices that can be stabilized by some cloud network control policy, given the cloud network structure $(\mathcal{G}, \Phi, \Delta)$. For each rate matrix $\boldsymbol{\lambda} \in \boldsymbol{\Lambda}_{\Delta}$, we denote by $h_{\Delta}^*(\boldsymbol{\lambda})$ the minimum average cost required for network stability.

The following theorem establishes that the capacity region and the minimum average cost for each arrival rate in the capacity region remains the same for any finite reconfiguration delay and cost. The proof of Theorem~\ref{theorem_capacity_region} is given in Appendix D.

\begin{theorem}
Given any finite reconfiguration delay/cost structure $\Delta$, the capacity region $\boldsymbol{\Lambda}_{\Delta}$ remains the same. In particular, $\boldsymbol{\Lambda}_{\Delta} = \boldsymbol{\Lambda}$, where $\boldsymbol{\Lambda}$ is the capacity region of the cloud network without reconfiguration delay, as characterized in~\cite[Theorem 1]{dcnc_info16}. 
Furthermore, given any exogenous arrival rate matrix $\boldsymbol{\lambda} \in \boldsymbol{\Lambda}_{\Delta}$, we have $h_{\Delta}^*(\boldsymbol{\lambda}) = h^*(\boldsymbol{\lambda})$.
\label{theorem_capacity_region}
\end{theorem}

While it was shown in~\cite{dcnc_icc16} that under the setup of no reconfiguration delay and cost, DCNC policy is throughput optimal and achieves a $[O(1/V), O(V)]$ cost-delay tradeoff, the result does not hold for the case in which reconfiguration delay or cost exists.
In fact, as it will be shown in section~\ref{Sec:simulation} (Figs.~\ref{queue_rate} and~\ref{cost_queue_reconfiguration_cost}), DCNC policy loses throughput optimality and the ability to achieve minimum average cost under the presence of reconfiguration delay or cost.

In the next section, we propose Adaptive DCNC (ADCNC) policy, which is an online distributed policy for cloud network control under reconfiguration delay and cost. We then establish theoretical performance guarantees of ADCNC policy, specifically throughput optimality and the $[O(1/V), O(V)]$ cost-delay tradeoff. In other words, ADCNC policy recovers the performance guarantees that DCNC policy loses when reconfiguration overhead exists.

\section{Dynamic Cloud Network Control under Reconfiguration Delay/Cost}\label{Sec:policy}

\subsection{Adaptive DCNC policy}

At each time slot $t$, each cloud network node makes local processing and transmission decisions on its corresponding outgoing interfaces. 

We select a function $g:\bbbr^+ \rightarrow \bbbr^+$ with $g(0) = 0$ which is strictly increasing and sublinear (i.e. $\lim\limits_{x \rightarrow\infty} \frac{g(x)}{x} = 0$), and a parameter $V \in \bbbr^+$. 
Given function $g$ and parameter $V$, node $i\in\mathcal V$ makes the following decisions at time $t$:

\begin{itemize}
\item \textbf{Transmission decisions:}
For each neighbor $j \in \mathcal V^+(i)$

\begin{enumerate}
\item Compute the {\em transmission max-utility-weight} as
\begin{align}
W_{ij}^*(t) = \max_{\substack{c \in \mathcal{C} \\ k \in \mathcal{K}_{ij}}} 
&\bigg\{ C_{ij}(k) \Big[ Q_{i}^{(c)}(t) - Q_{j}^{(c)}(t) - Ve_{ij} \Big]^+  \nonumber  \\
&\qquad  - Vw_{ij}(k) \bigg\}
\label{maxweight_def}
\end{align}
with $c^*$, $k^*$ being its maximizers. 

\item Let $(\bar{c}, \bar{k})$ denote the schedule at time $t-1$. Compute the {\em transmission weight} as
\begin{align}
W_{ij}(t) 
=& C_{ij}(\bar{k}) \left[ Q_{i}^{(\bar{c})}(t) - Q_{j}^{(\bar{c})}(t) - Ve_{ij} \right]^+  \nonumber \\
&- Vw_{ij}(\bar{k})
\end{align}

and the {\em transmission weight differential} at time $t$ as
\begin{align}
\Delta W_{ij}(t) 
=& W_{ij}^*(t) - W_{ij}(t)
\label{deltaW}
\end{align}

\item Define the {\em transmission weight differential threshold} at time $t$ as
\begin{align}
\theta_{ij}(t) = g\Big( C_{ij}(\bar{k}) (Q_{i}^{(c^*)}(t) - Q_{j}^{(c^*)}(t)) \Big)
\end{align}
and determine the transmission resource-commodity schedule at time $t$ as 

\begin{align*}
(k(t), c(t)) = 
\left\{
\begin{array}{ll}
(k^*, c^*) & \mbox{if } \Delta W_{ij}(t) > \theta_{ij}(t) \\
(\bar{k}, \bar{c}) & \mbox{otherwise}
\end{array}
\right.
\end{align*}

\item Allocate $k(t)$ transmission resource units and set transmission flow rates as:
\begin{flalign*}
&\mu_{ij}^{(c)}(t) = C_{ij}(k(t))\mathds{1}_{\{c = c(t)\}}, \forall c \in \mathcal{C}
\end{flalign*}

\end{enumerate}

\item \textbf{Processing decisions:} 

\begin{enumerate}
\item Compute the {\em processing max-utility weight} as
\begin{align}
W_{i}^*(t) = \max_{ \substack{c \in \mathcal{C} \\ k \in \mathcal{K}_{i}} } \bigg\{ 
& \frac{C_{i}(k)}{\rho_{i}^{(c)}} \Big[ Q_{i}^{(c)}(t) - \xi^{(c^+)} Q_{i}^{(c^+)}(t)    \nonumber \\
&\qquad - Ve_{i} \Big]^+   - Vw_{i}(k) \bigg\}
\end{align}  
with $c^*$, $k^*$ being its maximizers. 

\item Let $(\bar{c}, \bar{k})$ denote the schedule at time $t-1$. Compute the {\em processing weight} as
\begin{align}
W_{i}(t) 
=& \frac{C_{i}(\bar{k})}{\rho_{i}^{(\bar{c})}} \left[ Q_{i}^{(\bar{c})}(t) - \xi^{(\bar{c}^+)} Q_{i}^{(\bar{c}^+)}(t) - Ve_{i} \right]^+ \nonumber \\
& - Vw_{i}(\bar{k})
\end{align}

and the {\em processing weight differential} at time $t$ as
\begin{align}
\Delta W_{i}(t) 
=& W_{i}^*(t) - W_{i}(t)
\end{align}

\item Define the {\em processing weight differential threshold} at time $t$ as
\begin{align}
\theta_{i}(t) = g\Big( C_{i}(\bar{k}) (Q_{i}^{(c^*)}(t) -  Q_{i}^{(c^{*^+})}(t)) \Big)
\end{align}
and determine the processing resource-commodity schedule at time $t$ as 
\begin{align*}
(k(t), c(t)) = 
\left\{
\begin{array}{ll}
(k^*, c^*) & \mbox{if } \Delta W_{i}(t) > \theta_{i}(t) \\
(\bar{k}, \bar{c}) & \mbox{otherwise}
\end{array}
\right.
\end{align*}

\item Allocate $k(t)$ processing resource units and set processing flow rates as:
\begin{flalign*}
&\mu_{i}^{(c)}(t) = C_{i}(k(t))\mathds{1}_{\{c = c(t)\}}, \forall c \in \mathcal{C}
\end{flalign*}

\end{enumerate}

\end{itemize}

\subsection{Performance Analysis}
In this subsection, we extend the drift-plus-penalty analysis of~\cite{drift_plus_penalty} to show that Adaptive DCNC is throughput-optimal and achieves $[O(1/V), O(V)]$ average cost-delay tradeoff with probability 1 (w.p. 1) under any finite reconfiguration delay/cost.

The stability of Adaptive DCNC relies on the fact that it allows each node and link to adjust the frequency of reconfiguration according to its maximal queue length differential. In particular, Adaptive DCNC decreases the frequency of reconfiguration if the maximal queue length differential increases. This behavior may be characterized by the following lemma.

\begin{lemma}
Suppose Assumption~\ref{assumption_link}~and~\ref{assumption_arrivals} hold, and the cloud network is operated under Adaptive DCNC with parameter $V$ and sublinear function $g$. Given any fixed integer $T$, if the maximal queue length differential at a link $(i,j) \in \mathcal{E}$ at time $t$, $\max\limits_{c} \left\{Q_{i}^{(c)}(t) - Q_{j}^{(c)}(t) \right\}$, is greater than a constant $M_{ij}$ 
as defined below in~(\ref{Mij_def})
, then 
link $(i,j)$ reconfigures at most once during $[t, t+T]$. 

Similarly, if the maximal queue length differential at a node $i \in \mathcal{V}$ at time $t$, $\max\limits_{c} \left\{Q_{i}^{(c)}(t) - \xi^{(c^+)} Q_{i}^{(c^+)}(t) \right\}$, is greater than a constant $M_{i}$ as defined below in~(\ref{Mi_def})
, then 
node $i$ reconfigures at most once during $[t, t+T]$. 
\begin{align}
M_{ij} = 
\max \Big\{& V \Big( \min_{k>0}\frac{w_{ij}(k)}{C_{ij}(k)} + e_{ij} \Big) + T\gamma_{\max}, \nonumber \\
&\frac{1}{C_{ij}(1)} g^{-1}\left( 2C_{ij,\max} T \gamma_{\max} \right) + T\gamma_{\max} \Big\}
\label{Mij_def}  \\
M_{i} = 
\max \Big\{& V \Big( \min_{k>0}\frac{w_{ij}(k)}{C_{ij}(k)} + e_{ij} \Big) + T\gamma_{\max} , \nonumber \\
& \frac{1}{C_{i}(1)} g^{-1}\left( 2C_{i,\max} T \gamma_{\max} \right) +  T\gamma_{\max}  \Big\}
\label{Mi_def}
\end{align}
where $C_{ij,\max} = \max\limits_{k} C_{ij}(k)$, $C_{i,\max} = \max\limits_{k} C_{i}(k)$,  $\gamma_{\max} = \\ 2a_{\max} + 2C_{\max}(v_{\max}+1)$, $v_{\max} = \max\limits_{i} \{ \max\{|\mathcal{V}^+(i)|, \\ |\mathcal{V}^-(i)|\}\}$, and $C_{\max} = \max \{ \max\limits_{(i,j)\in\mathcal{E}}C_{ij,\max}, \max\limits_{i \in \mathcal{V}}C_{i,\max}\}$.

\label{lemma_reconfiguration}
\end{lemma}

The proof of Lemma~\ref{lemma_reconfiguration} is given in Appendix A. 

With Lemma~\ref{lemma_reconfiguration} limiting the frequency of reconfiguration, and the weight differentials $\{\Delta W_{i}(t)\}_{i \in \mathcal{V}}$, $\{\Delta W_{ij}(t)\}_{(i,j)\in\mathcal{E}}$ being bounded by local thresholds that are growing sublinearly with the local maximal queue length differential, we then extend the drift-plus-penalty analysis of~\cite{drift_plus_penalty} to prove the following performance guarantee. \\

\begin{theorem}
Suppose the arrival rate matrix $\boldsymbol{\lambda} = (\lambda_i^{(c)})$  is strictly interior to the capacity region $\boldsymbol{\Lambda}$, and suppose all reconfiguration delays and costs in $\Delta$ are finite. Then Adaptive DCNC stabilizes the cloud network, while achieving arbitrarily close to minimum average cost $h^*(\boldsymbol{\lambda})$ w.p. 1, i.e.
\vskip -0.2in
\begin{align}
&\lim\sup_{t \rightarrow \infty} \frac{1}{t} \sum_{\tau=0}^{t} h(\tau) \leq h^*(\boldsymbol{\lambda}) + \frac{B}{V} \label{eq8}\\
&\lim\sup_{t \rightarrow \infty} \frac{1}{t} \sum_{\tau=0}^{t} \sum_{\substack{i\in\mathcal{V}, \\ c\in \mathcal{C}}} Q_{i}^{(c)}(\tau)  \leq \frac{B + V[h^*(\boldsymbol{\lambda} + \epsilon \mathbf{1}) - h^*(\boldsymbol{\lambda})]}{\epsilon} \label{eq9}
\end{align}
where $B$ is a constant that is dependent on the system parameters $(\mathcal{G}, \Phi, \Delta), C_{i}(k), C_{ij}(k), w_{i}(k), w_{ij}(k), a_{\max}$; $\epsilon$ is a positive constant satisfying $(\boldsymbol{\lambda} + \epsilon \mathbf{1}) \in \boldsymbol{\Lambda}$, and $\mathbf{1}$ is a matrix of all ones.
\label{theorem_cost_delay}
\end{theorem}

The proof of Theorem~\ref{theorem_cost_delay} is given in Appendix B.

\section{Simulations} \label{Sec:simulation}

\begin{figure}[!t]
\centering
\includegraphics[height=2.1in]{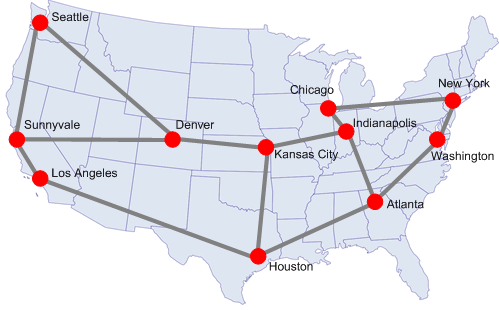}
\caption{Abilene US network topology. 
\vspace{-0.3in}
}
\label{abilene}
\end{figure}

In this section, we present simulation results for the proposed Adaptive DCNC policy and compare with benchmark policies. The sublinear function $g$ for Adaptive DCNC is selected as $g(x) = 0.99x^{0.99}$ for all simulations.

We consider a cloud network with network topology based on the Abilene US Network, as shown in Fig.~\ref{abilene}. We assume that all nodes are clouds that can host all service functions. We assume both nodes and links have homogeneous processing and transmission resources, respectively. 
In particular, we assume $K_i = 1$, $w_{i}(k) = k$ and $C_{i}(k) = k$ for each node $i\in\mathcal{V}$; while for each link $(i,j) \in \mathcal{E}$, we assume $K_{ij} = 1$, $w_{ij}(k) = k$, and $C_{ij}(k) = k$. 

We consider 2 services, each composed of 2 functions. We assume each service is requested by one source-destination pair. For Service 1, the source is in Seattle and the destination in New York; while for Service 2, the source is in Sunnyvale and the destination is in Atlanta. The arrival processe for both flows are i.i.d. Poisson with arrival rates denoted by $\lambda_1$ and $\lambda_2$, respectively. Throughout the simulation, we set both arrival rates to the same value, denoted by $\lambda$, i.e. $\lambda_1 = \lambda_2 = \lambda$.

For ease of discussion, we separate cases where only reconfiguration delays exist and where only reconfiguration costs exist in the following subsections.

\subsection{Reconfiguration Delay}

We first consider the case of cloud networks with reconfiguration delay only, in other words, the reconfiguration costs are set to zero. In this subsection, the reconfiguration delay of all the processing and transmission resources to be the same value, denoted by $\delta_r$.

\begin{figure}[!t]
\centering
\includegraphics[width=0.8\linewidth]{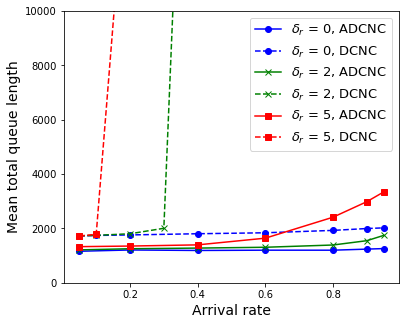}
\caption{Mean total queue length for DCNC and ADCNC under various flow arrival rates. Parameter $V = 0.5$.
\vspace{-0.3in}
}
\label{queue_rate}
\end{figure}

Fig.~\ref{queue_rate} compares the mean (time average) total queue length for DCNC and ADCNC under various flow arrival rates $\lambda$. The parameter $V$ is fixed as $V = 5.0$ for both policies. Given the topology and the processing and transmission capacity setting, the rate pair $(\lambda_1, \lambda_2) = (1.0, 1.0)$ is at the boundary of the capacity region, hence we consider the interval $\lambda \in [0, 1.0)$.
It is clear from the figure that when the reconfiguration delay $\delta_r$ is nonzero, DCNC loses throughput-optimality, and the maximum arrival rate it can stabilize reduces as the  reconfiguration delay $\delta_r$ increases. Note that while ADCNC shows smaller mean queue length than DCNC even when $\delta_r = 0$, ADCNC incurs slightly higher resource cost in this setting. Cost-delay tradeoff comparisons for input rates that both algorithms can stabilize are presented next.

\begin{figure}[!t]
\centering
\includegraphics[width=0.8\linewidth]{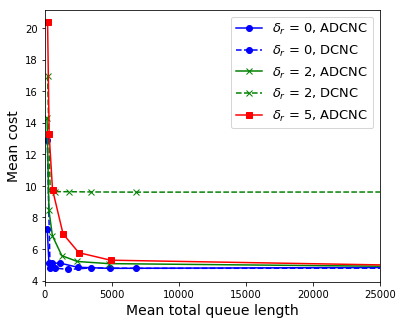}
\caption{Mean cost versus mean queue length for DCNC and ADCNC under various reconfiguration delays. Arrival rate $\lambda = 0.2$.
\vspace{-0.3in}
}
\label{cost_delay_reconfiguration_delay}
\end{figure}

In Fig.~\ref{cost_delay_reconfiguration_delay}, we plot the mean (time average) network cost versus the mean total queue length for DCNC and ADCNC under various reconfiguration delays.  
The arrival rate is fixed as $\lambda = 0.2$. Note that for each curve, the control parameter $V$ tunes the tradeoff between  network cost and  total queue length. The closer a curve  is  to the lower-left corner, the better the performance (cost-delay tradeoff). Note that without  reconfiguration delay, $\delta_r = 0$, DCNC and ADCNC have the similar performance. As $\delta_r$ increases, the performance of the two policies starts to degrade. Nevertheless, ADCNC policy always guarantees throughput-optimality, and is able to push the mean network cost arbitrarily close to minimum at the expense of increased mean queue length. In contrast, DCNC has significantly larger performance degradation as $\delta_r$ increases, and does not guarantee throughput-optimality. In fact, for $\delta_r = 5$, DCNC cannot even stabilize the arrival rate of $\lambda = 0.2$, hence the absence of the associated cost-delay curve.

\begin{figure}[!t]
\centering
\includegraphics[width=0.8\linewidth]{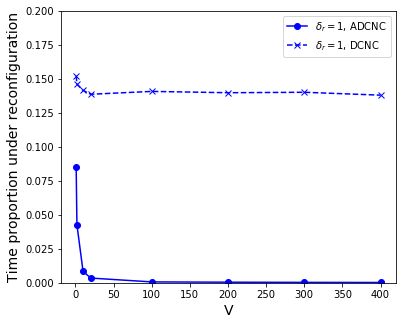}
\caption{Mean fraction of time under reconfiguration for various parameter V. 
\vspace{-0.3in}
}
\label{reconfig_V}
\end{figure}

\begin{figure}[!t]
\centering
\includegraphics[width=0.88\linewidth]{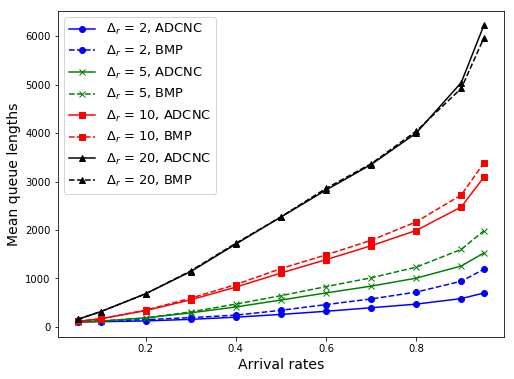}
\caption{Mean total queue length for ADCNC and BMP under various flow arrival rates. Sublinear function $g(x) = 0.99 x^{0.99}$ for both policies, and parameter $V = 0$ for ADCNC.
\vspace{-0.1in}
}
\label{ADCNC_BMP_compare}
\end{figure}

In Fig.~\ref{reconfig_V}, we further look into the reconfiguration behavior of both DCNC and ADCNC policy under various values of the control parameter $V$, with reconfiguration delay fixed to $\delta_r = 1$, and arrival rate $\lambda = 0.2$. The vertical axis represents the fraction of time that a given transmission/processing resource is under reconfiguration, averaged over all resources, i.e., the time overhead caused by the reconfiguration delay. We first notice that ADCNC spends much less time under reconfiguration, which is one of the key reasons for ADCNC to preserve throughput-optimality under finite reconfiguration delay. We then notice that while increasing the parameter $V$ helps reducing reconfiguration overhead for both policies, DCNC spends a significanlty higher fraction of time under reconfiguration even for large $V$.

To close the discussion of the reconfiguration delay case, we consider the comparison between ADCNC policy and Biased MaxPressure (BMP) policy proposed in~\cite{BMP}. Recall from Section~\ref{Sec:related} that BMP policy does not consider the cost minimization, which could be considered as a special case of the cloud network model in this paper. In this special case, we could set the parameter $V = 0$ for ADCNC policy in order to ignore the cost minimization, and make a fair comparison to BMP policy. In Fig.~\ref{ADCNC_BMP_compare}, we show the delay performance of ADCNC and BMP policies under varying arrival rates and under different reconfiguration delay. Since both policies are guaranteed to be throughput optimal, we could see that the delay remains finite for arrival rates $\lambda$ up to $1$. We could also see that while both policies have comparable delay performance, ADCNC performs slightly better especially for small reconfiguration delay.

\begin{figure}[!t]
\centering
\includegraphics[width=0.8\linewidth]{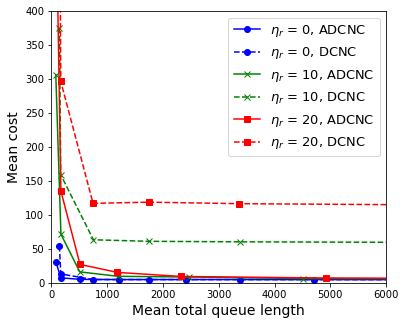}
\caption{Mean cost versus mean queue length for DCNC and ADCNC under various reconfiguration costs.
\vspace{-0.2in}
}
\label{cost_queue_reconfiguration_cost}
\end{figure}

\subsection{Reconfiguration Cost}

In this subsection, we set the reconfiguration delay to be zero, and set the reconfiguration cost of all the processing and transmission resource to be the same value, denoted by $\eta_r$.
Since there is no reconfiguration delay, both DCNC and ADCNC are throughput-optimal and can support the same arrival rates. 
We hence focus our attention in comparing their cost-delay tradeoff performance.

Fig.~\ref{cost_queue_reconfiguration_cost} shows the cost-delay tradeoff achieved by DCNC and ADCNC under various reconfiguration costs $\eta_r$.  We first notice that as the reconfiguration cost increases, DCNC can no longer achieve arbitrarily close to the minimum average cost, even when the parameter $V$ is tuned to endure large mean total queue length. 
On the other hand, Adaptive DCNC is able to achieve arbitrarily close to the minimum cost under any reconfiguration cost $\eta_r$.

\section{Extensions} \label{Sec:extension}

\subsection{Adaptive DCNC Policy for Generalized Setting}

In this subsection, we briefly discuss some interesting extensions to the current model that could be captured with slight modification of the analysis.

(1) \textsl{Different reconfiguration delay/cost for resource allocation and commodity allocation}: 
In this paper, we have assumed that the same reconfiguration delay and cost are incurred upon any change in either the allocation of resources or the commodity being processed/transmitted. In practice, different delays and costs can be associated with different reconfiguration operations. It is rather straightforward to show that ADCNC would preserve throughput and cost optimality for any finite values of such heterogeneous delays and costs by treating any change as incurring the maximum of such delays/costs. However, improved policies (in terms of cost-delay tradeoff) could be designed in such settings. In the next subsection, we provide an example heuristic variant of ADCNC policy to improve the cost-delay performance under this setting.

(2) \textsl{Partial reconfiguration:}
In this paper, we consider a worst-case reconfiguration delay model in the sense that we assume complete unavailability of packet processing or transmission functionality at a node or link undergoing reconfiguration. In practice, there may be cases in which adding or removing resources without changing the allocated commodity only reduces the available processing or transmission rate to the minimum between the available rates before and after reconfiguration. 
Importantly, a throughput-optimal policy for this worst-case reconfiguration delay model will guarantee throughput optimality for any other less restrictive model. 
Improved policies (in terms of cost-delay tradeoff) for this setting are of interest for future work.

\subsection{A Heuristic Variant of Adaptive DCNC Policy}

In the previous subsection, we introduced a more generic setting of cloud network where different reconfiguration overheads are associated with different reconfiguration operations, i.e. resource reconfiguration and commodity reconfiguration. While the throughput optimality guarantee for the Adaptive DCNC extends to this case as mentioned earlier, it is possible to improve the performance, i.e. cost-delay tradeoff, by exploiting the unequal reconfiguration overhead. We now introduce a heuristic variant of Adaptive DCNC policy as an example for improving the cost-delay performance.

Recall that Adaptive DCNC policy reconfigures both resource allocation and scheduled commodity at the same time. This approach is reasonable when the reconfiguration overhead is the same for both operations, since when one reconfiguration operation is performed, the other could be performed at the same time without incurring additional overhead. However, when the reconfiguration overhead is different for different operations, intuitively one may benefit from performing the reconfigure operation with smaller overhead more frequently. For this reason, we modify the reconfiguration criterion in Adaptive DCNC to a two-stage criterion. The first stage is the same as the reconfiguration criterion as in Adaptive DCNC, while the additional stage is used to decide whether to perform the reconfiguration operation with smaller overhead. With the additional stage of reconfiguration criterion, we could expect the reconfiguration operation with smaller overhead to be performed more frequently.

\begin{figure}[!t]
\centering
\includegraphics[width=0.89\linewidth]{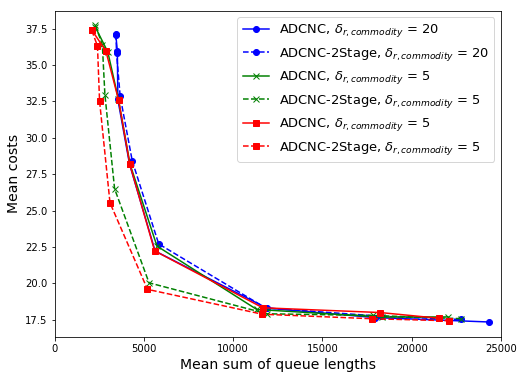}
\caption{Mean cost versus mean queue length for ADCNC and ADCNC-2stage under various commodity reconfiguration delay. Resource reconfiguration delay is fixed as $\delta_{r, resource} = 20$.
\vspace{-0.1in}
}
\label{cost_queue_separate_reconfiguration_delay}
\end{figure}

To be more specific, consider an example where the resource reconfiguration overhead is larger than the commodity reconfiguration overhead. For each cloud network node $i \in \mathcal{V}$, for the processing decision at each time instance $t$, node $i$ first follows steps 1) and 2) as described in section IV.A to compute $W_{i}^*(t)$ and $\Delta W_{i}(t)$. Then at step 3), node $i$ first checks if the criterion $\Delta W_{i}(t) > g(W_{i}^*(t))$ is met. If so, then reconfigure both resource and commodity allocation; otherwise, it further checks the following. Node $i$ computes $\Delta Q_{i}(t) = [Q_{i}^{(c^*)}(t) - Q_{i}^{({c^*}^+)}(t)]^+ - [Q_{i}^{(\bar{c})}(t) - Q_{i}^{(\bar{c}^+)}(t)]^+$, and compares it with a threshold $g([Q_{i}^{(c^*)}(t) - Q_{i}^{({c^*}^+)}(t)]^+)$. If $\Delta Q_{i}(t)$ is above the threshold, then reconfigures the commodity (while the resource allocation remains the same), otherwise no reconfiguration is performed. We refer this policy as ADCNC-2stage policy as it is a variant of ADCNC policy where the reconfiguration criterion becomes a 2 stage decision.

In Fig.~\ref{cost_queue_separate_reconfiguration_delay}, we show the simulation result for ADCNC and ADCNC-2stage under different commodity reconfiguration delay, while the resource reconfiguration delay is fixed as $\delta_{r, resource} = 20$. Again for simplicity we set all the reconfiguration cost to be zero. We first note that the performance of ADCNC policy (solid lines) remains similar. This aligns with the interpretation that ADCNC treats the reconfiguration overhead to be the maximum of the two different overheads. On the other hand, we could see that ADCNC-2stage policy (dashed lines) exploits the smaller commodity reconfiguration delay and improves the cost-delay performance as the commodity reconfiguration delay becomes smaller.

\section{Conclusion} \label{Sec:conclusion}

This paper addressed the dynamic control of network service chains in cloud networks with non-negligible resource reconfiguration delay and cost.
We showed that while the capacity region and the minimum achievable time average cost remains unchanged regardless of the value of reconfiguration delay or cost, the throughput  and cost optimality of existing policies (in the regime without reconfiguration delay/cost) is compromised when reconfiguration delay/cost exists. We then proposed Adaptive DCNC, a distributed flow scheduling and resource allocation policy that controls the frequency of reconfiguration based only on local queue length information. We showed that ADCNC is throughput optimal and achieves a  $[O(1/V), O(V)]$ cost-delay tradeoff, and validated the result via numerical simulations.

\bibliographystyle{ieeetr}
\bibliography{reference}

\appendix
In the following, given a time $t'$, we denote by
$({k}_{ij}(t'), {c}_{ij}(t'))$ the transmission resource-commodity pair scheduled at time $t'$ on link $(i,j)$, and by $({k}_{i}(t'), {c}_{i}(t'))$ the processing resource-commodity pair scheduled at time $t'$ at node $i$.
In addition,  we denote by $k_{ij}^*(t'),c_{ij}^*(t'))$ the transmission resource-commodity pair that maximizes the weight, $W^*_{ij}(t')$ at time $t'$ on link $(i,j)$, and by $(k_{i}^*(t'), c_{i}^*(t'))$ the processing resource-commodity pair that maximizes the weight, $W^*_{i}(t')$, at time $t'$ at node $i$.

\subsection{Proof of Lemma~\ref{lemma_reconfiguration}}

\begin{proof}
We prove the result by contradiction. Suppose that under the assumption of Lemma~\ref{lemma_reconfiguration}, there are two or more reconfigurations within the time period $[t, t+T]$. Therefore we may select two consecutive reconfiguration instances $t_1, t_2$ with $t \leq t_1 < t_2 < t+T$. 

Before we proceed with the proof,
we state the following lemmas that will be handy in the following. The proofs of these lemmas are given in Appendix C.

\begin{lemma}
Given Assumption~\ref{assumption_arrivals}, for any commodities $c_1, c_2 \in \mathcal{C}$, any nodes $i,j \in\mathcal{V}$, and any $\xi < \xi_{\max}$, $\rho > \rho_{\min}$, the (weighted) queue length differential between $Q_{i}^{(c_1)}$ and $Q_{j}^{(c_2)}$ can change only by a finite amount over one time slot, given as
\begin{align}
&\Big| \frac{1}{\rho} \Big(Q_{i}^{(c_1)}(\tau+1) - \xi Q_{j}^{(c_2)}(\tau+1) \Big)   \nonumber \\
& \qquad \qquad \qquad \qquad \qquad  - \frac{1}{\rho} \Big(Q_{i}^{(c_1)}(\tau) - \xi Q_{j}^{(c_2)}(\tau)\Big) \Big|  \nonumber \\
\leq& \frac{1}{\rho_{\min}} (1+\xi_{\max}) \big( a_{\max} + C_{\max} (v_{\max}+1) \big) \stackrel{\Delta}{=} \gamma_{\max}
\end{align}
where $C_{\max} = \max \{ \max\limits_{(i,j) \in \mathcal{E}} C_{ij}(K_{ij}), \max\limits_{i \in \mathcal{V}} C_{i}(K_{i}) \}$ is the maximum transmission or processing rate, and $v_{\max} = \max\limits_{i} \{ \max\{|\mathcal{V}^+(i)|, |\mathcal{V}^-(i)|\}\}$ is the maximum number of incoming or outgoing links over all nodes. We also take $\rho_{\min} = \min\{1, \min\limits_{i,c} \rho_{i}^{(c)}\}$ and $\xi_{\max} = \max\{1, \max\limits_{c} \xi^{(c)}\}$.

Similarly, the change in the maximal queue length differential for transmission on link $(i,j)$ over one time slot is bounded as
\begin{align}
\Big| \max_{c \in \mathcal{C}} \Big\{Q_{i}^{(c)}(\tau+1) &- Q_{j}^{(c)}(\tau+1) \Big\} \nonumber \\ 
&- \max_{c \in \mathcal{C}} \Big\{Q_{i}^{(c)}(\tau) - Q_{j}^{(c)}(\tau)\Big\} \Big| 
\leq \gamma_{\max}
\label{eq:max_diff_tx_bound}
\end{align}
and the change in the maximal queue length differential for processing in node $i$ over one time slot is bounded as
\begin{align}
\bigg|& \max_{c \in \mathcal{C}} \Big\{ \frac{1}{\rho_{i}^{(c)}} \big[ Q_{i}^{(c)}(\tau+1) - \xi^{(c^+)} Q_{i}^{(c^+)}(\tau+1) - Ve_{i} \big]^+ \Big\} \nonumber \\ 
&- \max_{c \in \mathcal{C}} \Big\{ \frac{1}{\rho_{i}^{(c)}} \big[ Q_{i}^{(c)}(\tau) - \xi^{(c^+)} Q_{i}^{(c^+)}(\tau) -Ve_{i} \big]^+\Big\} \bigg| 
\leq \gamma_{\max}
\label{eq:max_diff_px_bound}
\end{align}

\label{lemma_change_bound}
\end{lemma}

\begin{lemma}
Given Assumption~\ref{assumption_link}, and any fixed $V < \infty$, define $F(x) \stackrel{\Delta}{=} \max\limits_{k \leq K_{ij}} \big\{ C_{ij}(k) [x - Ve_{ij}]^+ - Vw_{ij}(k) \big\}$. Then, 
\begin{enumerate}
\item[(a)] $F(x)$ is Lipschitz continuous with Lipschitz constant $C_{ij,\max} = \max\limits_{k \leq K_{ij}}C_{ij}(k)$. 

\item[(b)] If $x > V \big( \min\limits_{k>0}\frac{ w_{ij}(k)}{ C_{ij}(k)} + e_{ij} \big)$, then $F(x) > 0$ and the $k^*$ maximizing $F(x)$ satisfies $k^* > 0$; otherwise, $F(x) = 0$ and the maximizer is $k^* = 0$.

\end{enumerate}
\label{lemma_resource_maximization}
\end{lemma}



In the following, we show that under the assumption Lemma~\ref{lemma_reconfiguration},
$\max\limits_{c} \left\{Q_{i}^{(c)}(t) - Q_{j}^{(c)}(t) \right\} > M_{ij}$, the weight difference at time $t_2$ can not exceed the threshold $g\Big( C_{ij}({k}_{ij}(t_2-1)) \max\limits_{c \in \mathcal{C}} \left\{Q_{i}^{(c)}(t_2) - Q_{j}^{(c)}(t_2) \right\}  \Big)$.
This, hence, contradicts  the assumption that Adptive DCNC reconfigures at time $t_2$.

To do this, starting from \eqref{deltaW}
we rewrite the transmission weight differential as:  
\begin{align*}
\Delta W_{ij}(t_2) 
=&  W_{ij}^*(t_2) - W_{ij}(t_2) \nonumber \\
=&  (W_{ij}^*(t_2) - W_{ij}^*(t_1) ) + (W_{ij}^*(t_1)- W_{ij}(t_2))
\end{align*}
with
\begin{align*}
W_{ij}^*(t_2)
=& F\Big(\max_{c \in \mathcal{C}} \left\{Q_{i}^{(c)}(t_2) - Q_{j}^{(c)}(t_2) \right\}\Big)
\nonumber \\
=&
C_{ij}(k_{ij}^*(t_2)) \Big[Q_{i}^{(c_{ij}^*(t_2))}(t_2) - Q_{j}^{(c_{ij}^*(t_2))}(t_2) -Ve_{ij} \Big]^+ \nonumber\\
&- Vw_{ij}(k_{ij}^*(t_2)) 
\end{align*}
\begin{align*}
W_{ij}^*(t_1)
=&
F\Big(\max_{c \in \mathcal{C}} \left\{Q_{i}^{(c)}(t_1) - Q_{j}^{(c)}(t_1) \right\}\Big)
\nonumber\\
=&C_{ij}(k_{ij}^*(t_1)) \Big [Q_{i}^{(c_{ij}^*(t_1))}(t_1) - Q_{j}^{(c_{ij}^*(t_1))}(t_1) -Ve_{ij} \Big]^+    \nonumber\\
&- Vw_{ij}(k_{ij}^*(t_1)) 
\end{align*}
and 
\begin{align}
W_{ij}(t_2)=&  
C_{ij}(\bar{k}_{ij}) \Big [Q_{i}^{(\bar{c}_{ij})}(t_2) - Q_{j}^{(\bar{c}_{ij})}(t_2) -Ve_{ij} \Big]^+ \nonumber
\\  & \quad   - Vw_{ij}(\bar{k}_{ij}) \nonumber \\
=& C_{ij}(k_{ij}^*(t_1)) \Big [Q_{i}^{(c_{ij}^*(t_1))}(t_2) - Q_{j}^{(c_{ij}^*(t_1))}(t_2) -Ve_{ij} \Big]^+ \nonumber
\\ & - Vw_{ij}(k_{ij}^*(t_1))
\label{eq:final}
\end{align}
where in \eqref{eq:final} we have used the fact that, given the  assumption that Adaptive DCNC reconfigures at time slots $t_1$, $t_2$ 
, during $[t_1, t_2-1]$ the resource allocation and the transmitted commodity remains $k_{ij}^*(t_1)$ and $c_{ij}^*(t_1)$, respectively.


Now, using Lemma~\ref{lemma_resource_maximization}~(a), we have
\begin{align}
&W_{ij}^*(t_2) - W_{ij}^*(t_1)  \nonumber \\
=& F\Big(\max_{c \in \mathcal{C}} \left\{Q_{i}^{(c)}(t_2) - Q_{j}^{(c)}(t_2) \right\}\Big)   \nonumber \\
& \qquad \qquad - F\Big(\max_{c \in \mathcal{C}} \left\{Q_{i}^{(c)}(t_1) - Q_{j}^{(c)}(t_1) \right\}\Big)  \nonumber \\
\leq& C_{ij,\max} \Big| \max\limits_{c \in \mathcal{C}}\left\{Q_{i}^{(c)}(t_2) - Q_{j}^{(c)}(t_2) \right\}  \nonumber \\
& \qquad \qquad  - \max\limits_{c \in \mathcal{C}}\left\{Q_{i}^{(c)}(t_1) - Q_{j}^{(c)}(t_1) \right\} \Big|  \nonumber \\
\leq& C_{ij,\max} (t_2 - t_1) \gamma_{\max} \leq C_{ij,\max} T \gamma_{\max}
\label{maxweight_change_bound}
\end{align}

%

On the other hand we have that:
\begin{align}
&W_{ij}^*(t_1) - W_{ij}(t_2)  \nonumber \\
=& C_{ij}(k_{ij}^*(t_1)) [Q_{i}^{(c_{ij}^*(t_1))}(t_1) - Q_{j}^{(c_{ij}^*(t_1))}(t_1) -Ve_{ij}]^+  \nonumber \\
&- C_{ij}(k_{ij}^*(t_1)) [Q_{i}^{(c_{ij}^*(t_1))}(t_2) - Q_{j}^{(c_{ij}^*(t_1))}(t_2) -Ve_{ij}]^+  \nonumber \\
\leq& C_{ij}(k_{ij}^*(t_1)) \Big| \left(Q_{i}^{(c_{ij}^*(t_1))}(t_2) - Q_{j}^{(c_{ij}^*(t_1))}(t_2) \right)  \nonumber \\
& \ \ \ \ \ \ \ \ \ \ \ \ \ \ \ \ - \left(Q_{i}^{(c_{ij}^*(t_1))}(t_1) - Q_{j}^{(c_{ij}^*(t_1))}(t_1) \right) \Big|  \nonumber \\
\leq& C_{ij,\max} (t_2 - t_1) \gamma_{\max} \leq C_{ij,\max} T \gamma_{\max}
\label{weight_change_bound}
\end{align}
where the last inequality follows from Lemma~\ref{lemma_change_bound}.

Combining (\ref{maxweight_change_bound}) and (\ref{weight_change_bound}), we have 
\begin{align}
W_{ij}^*(t_2) - W_{ij}(t_2) \leq 2C_{ij,\max}T\gamma_{\max}
\label{weight_difference_bound}
\end{align}





From the assumption of Lemma~\ref{lemma_reconfiguration}, and using Lemma~\ref{lemma_change_bound}, we have $\max\limits_{c \in \mathcal{C}} \big\{Q_{i}^{(c)}(t_1) - Q_{j}^{(c)}(t_1)\big\} > V \big( \min\limits_{k>0}\frac{ w_{ij}(k)}{ C_{ij}(k)} + e_{ij} \big)$, and hence $k_{ij}^*(t_1) > 0$ following Lemma~\ref{lemma_resource_maximization}~(b). Similarly, with the assumption of 
Lemma~\ref{lemma_reconfiguration}, and using Lemma~\ref{lemma_change_bound}, we also have $\max\limits_{c \in \mathcal{C}} \big\{Q_{i}^{(c)}(t_2) - Q_{j}^{(c)}(t_2)\big\} > \frac{1}{C_{ij}(1)} g^{-1} \big( 2 C_{ij,\max} T\gamma_{\max} \big)$, 
from which it follows that 
\begin{align}
&g\Big( C_{ij}(k_{ij}^*(t_1)) \max\limits_{c \in \mathcal{C}}   \left\{Q_{i}^{(c)}(t_2) - Q_{j}^{(c)}(t_2) \right\}  \Big) \nonumber\\
>& 2 C_{ij,\max} T\gamma_{\max} \nonumber\\
\geq&  W_{ij}^*(t_2) - W_{ij}(t_2)
\label{contradiction}
\end{align}
which 
contradics the assumption that Adaptive DCNC reconfigures at time $t_2$.


We now consider the condition for the processing decision at node $i$. From the assumption of Lemma~\ref{lemma_reconfiguration}, $\max\limits_{c} \left\{Q_{i}^{(c)}(t) - \xi^{(c^+)} Q_{i}^{(c^+)}(t) \right\} > M_{i}$, 

To do this, 
we rewrite the processing weight differential as:  
\begin{align*}
\Delta W_{i}(t_2) 
=&  W_{i}^*(t_2) - W_{i}(t_2) \nonumber \\
=&  (W_{i}^*(t_2) - W_{i}^*(t_1) ) + (W_{i}^*(t_1)- W_{i}(t_2))
\end{align*}
with
\begin{align*}
W_{i}^*(t_2)
=& \frac{C_{i}(k_{i}^*(t_2))}{\rho_{i}^{(c^*_{i}(t_2))}} \Big[Q_{i}^{(c_{i}^*(t_2))}(t_2) - \xi^{(c_{i}^{*^+}(t_2))} Q_{i}^{(c_{i}^{*^+}(t_2))}(t_2) \nonumber\\
& -Ve_{i} \Big]^+  - Vw_{i}(k_{i}^*(t_2)) 
\end{align*}
\begin{align*}
W_{i}^*(t_1)
=& \frac{C_{i}(k_{i}^*(t_1))}{\rho_{i}^{(c^*_{i}(t_1))}} \Big[Q_{i}^{(c_{i}^*(t_1))}(t_1) - \xi^{(c_{i}^{*^+}(t_1))} Q_{i}^{(c_{i}^{*^+}(t_1))}(t_1) \nonumber\\
& -Ve_{i} \Big]^+  - Vw_{i}(k_{i}^*(t_1)) 
\end{align*}
and 
\begin{align}
W_{i}(t_2)
=& \frac{C_{i}(k_{i}^*(t_1))}{\rho_{i}^{(c_{i}^{*^+}(t_1))}} \Big [Q_{i}^{(c_{i}^*(t_1))}(t_2) - \xi^{(c_{i}^{*^+}(t_1))} Q_{i}^{(c_{i}^{*^+}(t_1))}(t_2)  \nonumber
\\ &-Ve_{i} \Big]^+ - Vw_{i}(k_{i}^*(t_1))
\label{eq:final2}
\end{align}
where in \eqref{eq:final2} we have used the fact that, given the  assumption that Adaptive DCNC reconfigures at time slots $t_1$, $t_2$, during $[t_1, t_2-1]$ the resource allocation and the commodity being processed remains $k_{i}^*(t_1), c_{i}^*(t_1)$, respectively.

Now, using Lemma~\ref{lemma_resource_maximization}~(a), we have
\begin{align}
&W_{i}^*(t_2) - W_{i}^*(t_1)  \nonumber \\
=& F\Big(\max_{c \in \mathcal{C}} \Big\{  \frac{1}{\rho_{i}^{(c)}} [Q_{i}^{(c)}(t_2) - Q_{j}^{(c)}(t_2) -Ve_{i}]^+ \Big\}\Big)   \nonumber \\
& \qquad \qquad - F\Big(\max_{c \in \mathcal{C}} \Big\{  \frac{1}{\rho_{i}^{(c)}} [Q_{i}^{(c)}(t_1) - Q_{j}^{(c)}(t_1) -Ve_{i}]^+ \Big\} \Big)  \nonumber \\
\leq& C_{ij,\max} \Big| \max\limits_{c \in \mathcal{C}}\Big\{  \frac{1}{\rho_{i}^{(c)}} [Q_{i}^{(c)}(t_2) - Q_{j}^{(c)}(t_2) -Ve_{i}]^+ \Big\} \nonumber \\
& \qquad \qquad  - \max\limits_{c \in \mathcal{C}}\Big\{  \frac{1}{\rho_{i}^{(c)}} [Q_{i}^{(c)}(t_1) - Q_{j}^{(c)}(t_1) -Ve_{i}]^+ \Big\} \Big|  \nonumber \\
\leq& C_{ij,\max} (t_2 - t_1) \gamma_{\max} \leq C_{ij,\max} T \gamma_{\max}
\label{maxweight_change_bound2}
\end{align}

\end{proof}

\subsection{Proof of Theorem~\ref{theorem_cost_delay}}
\begin{proof}
Consider the quadratic Lyapunov function $L:\bbbr^{N \times N} \rightarrow \bbbr$ where $L(\mathbf{Q}) = \frac{1}{2} \sum_{i,c} (Q_{i}^{(c)})^2$. Let $\mathbf{X}(t) = \Big( \mathbf{Q}(t), \mathbf{k}(t), \boldsymbol{\mu}(t), \mathbf{r}(t) \Big)$ denote the queue length, resource allocation decision, flow rate decision, and the reconfiguration status of the cloud network at time $t$.

We first consider zero reconfiguration costs $\eta_{i} = 0, \eta_{ij}=0$.

We now leverage Lemma~\ref{lemma_reconfiguration} to bound the $T$-step Lyapunov drift-plus-penalty.
\begin{align}
&\bbbe \bigg[ L(\mathbf{Q}(t+T)) - L(\mathbf{Q}(t)) + V\sum_{\tau=t}^{t+T} h(\tau) \bigg| \mathbf{X}(t) \bigg]  \nonumber \\
=& \bbbe \bigg[ \sum_{\tau=t}^{t+T-1} \bbbe \left[ L(\mathbf{Q}(\tau+1)) - L(\mathbf{Q}(\tau)) + V h(\tau) | \mathbf{X}(\tau) \right]  \bigg| \mathbf{X}(t) \bigg] 
\label{T_drift}
\end{align}
For each time slot $\tau \in [t, t+T]$ we have:
\begin{align}
&\bbbe \left[ L(\mathbf{Q}(\tau+1)) - L(\mathbf{Q}(\tau)) + V h(\tau) \bigg| \mathbf{X}(\tau) \right]   \nonumber \\
\leq& \Phi + \sum_{i,c} Q_{i}^{(c)}(\tau) \lambda_i^{(c)} - \bbbe \Big[ Z(\tau) \Big| \mathbf{X}(\tau) \Big] + V \bbbe\Big[h(\tau) \Big| \mathbf{X}(\tau) \Big] 
\label{drift_penalty}
\end{align}
where
\begin{align}
\Phi =& \frac{1}{2} N 
\Big[ ( (v_{\max}+1) C_{\max})^2 + ((v_{\max} + 1) C_{\max} + a_{\max})^2 \Big]  \nonumber \\
Z(\tau) &= \sum_{i,c} Q_{i}^{(c)}(\tau) \Big[ \sum_{j\in\delta(i)} \mu_{ij}^{(c)}(\tau) \mathds{1}_{\{r_{ij}(\tau)=0\}} \nonumber \\
& \qquad  + \mu_{i}^{(c)}(\tau)\mathds{1}_{\{r_{i}(\tau)=0\}}  - \sum_{j:i\in\delta(j)} \mu_{ji}^{(c)}(\tau) \mathds{1}_{\{r_{ji}(\tau)=0\}}  \nonumber \\
& \qquad \qquad- \mu_{i}^{(c^-)}(\tau)\mathds{1}_{\{r_{i}(\tau)=0\}} 
\Big]
\end{align}
and $v_{\max} = \max\limits_{i\in\mathcal{V}}\{ \max\{ |\mathcal{V}^+(i)|, |\mathcal{V}^-(i)|\} \}$.

Let  
\begin{align}
W^*(\tau) &= 
\sum_{(i,j)\in\mathcal{E}}
W_{ij}^*(\tau) + \sum_{i \in\mathcal{V}}  
W_{i}^*(\tau), \label{maxweight_total}
\end{align}

where the transmission max-utility-weights and the processing max-utility-weights are given by
\begin{align*}
W_{ij}^*(\tau) 
=& \mu_{ij}^{*^{(c_{ij}^*(\tau)})}(\tau) \big(Q_{i}^{(c_{ij}^*(\tau)}(\tau) - Q_{j}^{(c_{ij}^*(\tau))}(\tau) -Ve_{ij} \big) \\
&- Vw_{ij}(k_{ij}^*(\tau))  \\
W_{i}^*(\tau) 
=& \mu_{i}^{*^{(c_{i}^*(\tau))}}(\tau) \big(Q_{i}^{(c_{i}^*(\tau))}(\tau) -  Q_{i}^{(c_i^{*^+}(\tau))}(\tau) -Ve_{i} \big) \\
&- Vw_{i}(k_{i}^*(\tau)) 
\end{align*}




In \eqref{drift_penalty}, adding and subtracting  $ \bbbe  \Big[ W^*(\tau) \Big| \mathbf{X}(\tau) \Big]$ given in \eqref{maxweight_total}, and recalling that for any $\epsilon > 0$ such that $\boldsymbol{\lambda} + \epsilon \mathbf{1} \in \boldsymbol{\Lambda}$, (see \cite{dcnc_info16,dcnc_icc16} for derivation):
\begin{align}
&\sum_{i,c} Q_{i}^{(c)}(\tau) \lambda_i^{(c)} 
- \bbbe \Big[ W^*(\tau) \Big| \mathbf{X}(\tau) \Big]  \nonumber \\
\leq& -\epsilon \sum_{i,c} Q_{i}^{(c)}(\tau) + V h^*(\boldsymbol{\lambda} + \epsilon \mathbf{1}),
\label{negative_drift}
\end{align}
 we can further bound  the drift-plus-penalty in \eqref{drift_penalty} as:
\begin{align}
&\bbbe \left[ L(\mathbf{Q}(\tau+1)) - L(\mathbf{Q}(\tau)) + V h(\tau) \bigg| \mathbf{X}(\tau) \right]   \nonumber \\
\leq& \Phi  -\epsilon \sum_{i,c} Q_{i}^{(c)}(\tau) + V h^*(\boldsymbol{\lambda} + \epsilon \mathbf{1}) \nonumber \\
&+ \bbbe  \Big[ W^*(\tau) \Big| \mathbf{X}(\tau) \Big]
- \bbbe \Big[ Z(\tau) \Big| \mathbf{X}(\tau) \Big] + V \bbbe\Big[h(\tau) \Big| \mathbf{X}(\tau) \Big] 
\label{drift_penalty1}
\end{align}

From above we have a negative term in the drift-plus-penalty which decreases as the total queue length increases. It then remains to ensure that $W^*(\tau) - Z(\tau) + Vh(\tau)$ could be bounded so that the we can still bound the drift-plus-penalty with a term that decrease when the 
total queue length increases.
To this end notice that: 

\begin{align}
&W^*(\tau) - Z(\tau) + Vh(\tau)  \nonumber \\
=& \sum_{(i,j)\in\mathcal{E}} \bigg[  
W_{ij}^*(\tau) - W_{ij}(\tau) \mathds{1}_{\{r_{ij}(\tau)=0\}}
\bigg]  \nonumber \\
&+ \sum_{i \in\mathcal{V}} \Big[  
W_{i}^*(\tau) - W_{i}(\tau)\mathds{1}_{\{r_{i}(\tau)=0\}}
\Big] 
\label{maxweight_current_weight_diff}
\end{align}
where 
\begin{align*}
W_{ij}(\tau) 
=&  \mu_{ij}^{(c(\tau))}(\tau) \big(Q_{i}^{(c(\tau))}(\tau) - Q_{j}^{(c(\tau))}(\tau) -Ve_{ij} \big)  \\
&- Vw_{ij}(k_{ij}(\tau))  \\
W_{i}(\tau) 
=& \mu_{i}^{(c(\tau))}(\tau) \big(Q_{i}^{(c(\tau))}(\tau) - 
Q_{i}^{(c^+(\tau))}(\tau) -Ve_{i} \big)  \\
&- Vw_{i}(k_{i}(\tau)) 
\end{align*}

For the above expression, we now bound the term for each node $i$ and each link $(i,j)$ separately. We do this with the help of Lemma~\ref{lemma_reconfiguration}.  

We start with the term for link $(i,j)$. 
Note that for any given  time $\tau \in [t, t+T]$ such that Adaptive DCNC does not reconfigure (i.e $r_{ij}(\tau) = 0$), by  construction we have
that the transmission weight differential is bounded by:
\begin{align}
W_{ij}^*(\tau) - W_{ij}(\tau) 
\leq& g \bigg( C_{ij,\max} \max_{c}\Big\{ Q_{i}^{(c)}(\tau) - Q_{j}^{(c)}(\tau) \Big\} \bigg)  \nonumber \\
\leq& g \bigg( C_{\max} \max_{i,c}\Big\{ Q_{i}^{(c)}(\tau) \Big\} \bigg).
\label{weight_difference_1}
\end{align}
Alternative for any given  time $\tau \in [t, t+T]$ such that  $r_{ij}(\tau) > 0$ i.e. Adaptive DCNC is under reconfiguration, since the transmission weight differential
is always bounded by the transmission max-utility-weight, than it suffices to bound  $W_{ij}^*(\tau)$. In particular, if link $(i,j)$ at time $t$ satisfies $\max\limits_{c} \{Q_{i}^{(c)}(t) - Q_{j}^{(c)}(t) \} > M_{ij}$, then  at time $\tau$ the transmission weight differential can be bound as follow:
\begin{align}
W_{ij}^*(\tau) - W_{ij}(\tau)  &\leq W_{ij}^*(\tau) \nonumber \\
&\leq C_{ij,\max}  \max_{c}\Big\{ Q_{i}^{(c)}(\tau) - Q_{j}^{(c)}(\tau) \Big\}  \nonumber \\
&\leq C_{\max} \max_{i,c}\Big\{ Q_{i}^{(c)}(\tau) \Big\} 
\label{weight_difference_2}
\end{align}
while for the case that at time $t$, $\max\limits_{c} \{Q_{i}^{(c)}(t) - Q_{j}^{(c)}(t) \} \leq M_{ij}$, 
starting from the expression of  the transmission max-utility-weight $W_{ij}^*(\tau)$, we observe that since $M_{ij}$ is the max of two terms one of the following could hold:  
\begin{enumerate}
    \item $\max\limits_{c} \{Q_{i}^{(c)}(t) - Q_{j}^{(c)}(t) \} \leq V \Big( \min\limits_{k}\frac{w_{ij}(k)}{C_{ij}(k)} + e_{ij} \Big) + T\gamma_{\max}$. \\
    In this case using Lemma~\ref{lemma_change_bound},   we  have that at time $\tau$
    \begin{align}
    \max\limits_{c} \{Q_{i}^{(c)}(\tau) - Q_{j}^{(c)}(\tau) \} &\leq V \Big( \min\limits_{k}\frac{w_{ij}(k)}{C_{ij}(k)} + e_{ij} \Big) \nonumber \\
    & \qquad + 2T\gamma_{\max}
    \end{align}
    from which it follows that 
    \begin{align*}
    W_{ij}^*(\tau) 
    \leq& \max_{k} \Big\{ C_{ij}(k) \big[V\min\limits_{k}\frac{w_{ij}(k)}{C_{ij}(k)} + 2T\gamma_{\max} \big]  \nonumber \\
    &\ \ \ \ \ \ \ \ - Vw_{ij}(k) \Big\}  \nonumber \\
    \leq& \max_{k} \Big\{ C_{ij}(k)   V\min\limits_{k}\frac{w_{ij}(k)}{C_{ij}(k)} - Vw_{ij}(k) \Big\} \nonumber \\
    &+ 2C_{ij,\max} T\gamma_{\max}  \nonumber \\
    \leq& 2C_{ij,\max} T\gamma_{\max} 
    \end{align*}
    where the last inequality follows from  Lemma~\ref{lemma_resource_maximization}~(b).
    
    \item $\max\limits_{c} \{Q_{i}^{(c)}(t) - Q_{j}^{(c)}(t) \} \leq \frac{1}{C_{ij}(1)} g^{-1} \Big( 2 C_{ij,\max} T\gamma_{\max} \Big)\\ + T\gamma_{\max}$: \\
    In this case using Lemma~\ref{lemma_change_bound},   we  have that at time $\tau$
    \begin{align} 
    \max\limits_{c} \{Q_{i}^{(c)}(\tau) - Q_{j}^{(c)}(\tau) \} &\leq \frac{1}{C_{ij}(1)} g^{-1} \Big( 2 C_{ij,\max} T\gamma_{\max} \Big) \nonumber \\
    & \qquad + 2T\gamma_{\max}
    \end{align}
    and thus
    \begin{align*}
    W_{ij}^*(\tau)
    \leq& C_{ij,\max}  \max_{c}\Big\{ Q_{i}^{(c)}(\tau) - Q_{j}^{(c)}(\tau) \Big\}  \nonumber \\
    \leq& \frac{C_{ij,\max}}{C_{ij}(1)} g^{-1} \Big( 2 C_{ij,\max} T\gamma_{\max} \Big) \nonumber \\
    &+ 2C_{ij,\max}T\gamma_{\max} 
    \end{align*}
\end{enumerate}
Combining the two cases, we have for $\max\limits_{c} \{Q_{i}^{(c)}(t) - Q_{j}^{(c)}(t) \} \leq M_{ij}$ that
\begin{align}
W_{ij}^*(\tau) &\leq \frac{C_{ij,\max}}{C_{ij}(1)} g^{-1} \Big( 2 C_{ij,\max} T\gamma_{\max} \Big) + 2C_{ij,\max}T\gamma_{\max}  \nonumber \\
&\stackrel{\Delta}{=} \Phi_{ij}
\label{weight_difference_3}
\end{align}



We may then use the same approach to give a bound on the term for each node $i$.

Applying (\ref{drift_penalty}), (\ref{negative_drift}), (\ref{maxweight_current_weight_diff}), and (\ref{weight_difference_1})-(\ref{weight_difference_3}) into (\ref{T_drift}), we have

\begin{align}
&\bbbe \left[ L(\mathbf{Q}(t+T)) - L(\mathbf{Q}(t)) + V\sum_{\tau=t}^{t+T} h(\tau) | \mathbf{X}(t) \right]  \nonumber \\
\leq& \bbbe \bigg[ 
T\Phi - \epsilon \sum_{\tau=t}^{t+T-1} \sum_{i,c}Q_{i}^{(c)}(\tau) +TVh^*(\boldsymbol{\lambda} + \epsilon \mathbf{1})  \nonumber \\
&+ \sum_{\tau=t}^{t+T} \bigg( \sum_{(i,j)\in\mathcal{E}} g \Big( C_{\max} \max_{i,c}\Big\{ Q_{i}^{(c)}(\tau) \Big\} \Big) \mathds{1}_{\{r_{ij}(\tau)=0\}}  \nonumber \\
&+  \sum_{ \substack{(i,j)\in\mathcal{E}: \\ \max\limits_{c} \{Q_{i}^{(c)}(t) -  Q_{j}^{(c)}(t) \} > M_{ij}} } 
C_{\max} \max_{i,c}\Big\{ Q_{i}^{(c)}(\tau) \Big\} \mathds{1}_{\{r_{ij}(\tau)>0\}} \nonumber \\
&+  \sum_{ \substack{(i,j)\in\mathcal{E}: \\ \max\limits_{c} \{Q_{i}^{(c)}(t) -  Q_{j}^{(c)}(t) \} \leq M_{ij}} } 
\Phi_{ij} \mathds{1}_{\{r_{ij}(\tau)>0\}} 
\bigg)  \nonumber \\
&+ \sum_{\tau=t}^{t+T} \bigg( \sum_{i\in\mathcal{V}} g \Big( C_{\max} \max_{i,c}\Big\{ Q_{i}^{(c)}(\tau) \Big\} \Big) \mathds{1}_{\{r_{i}(\tau)=0\}}  \nonumber \\
&+  \sum_{ \substack{i\in\mathcal{V}: \\ \max\limits_{c} \{Q_{i}^{(c)}(t) - Q_{i}^{(c^+)}(t) \} > M_{i}} } 
C_{\max} \max_{i,c}\Big\{ Q_{i}^{(c)}(\tau) \Big\} \mathds{1}_{\{r_{i}(\tau)>0\}} \nonumber \\
&+  \sum_{ \substack{i\in\mathcal{V}: \\ \max\limits_{c} \{Q_{i}^{(c)}(t) - Q_{i}^{(c^+)}(t) \} \leq M_{i}} } 
\Phi_{i} \mathds{1}_{\{r_{i}(\tau)>0\}} 
\bigg)  \nonumber 
%
%
\end{align}
\begin{align}
\leq& \bbbe \bigg[ 
T\Phi - \epsilon \sum_{\tau=t}^{t+T-1} \sum_{i,c}Q_{i}^{(c)}(\tau) +TVh^*(\boldsymbol{\lambda} + \epsilon \mathbf{1})  \nonumber \\
&+ \sum_{(i,j)\in\mathcal{E}} \Big[ \sum_{\tau=t}^{t+T} g\Big( C_{\max} \max_{i,c}\big\{ Q_{i}^{(c)}(\tau) \big\} \Big)   \nonumber \\
&\ \ \ \ \ \ \ 
+  \delta_{ij} C_{\max} \Big( \max_{i,c}\big\{ Q_{i}^{(c)}(t) \big\} + T\gamma_{\max} \Big) + T\Phi_{ij}  \Big] \nonumber \\
&+ \sum_{i\in\mathcal{V}} \Big[  \sum_{\tau=t}^{t+T} g\Big( C_{\max} \max_{i,c}\big\{ Q_{i}^{(c)}(\tau) \big\} \Big) \nonumber \\
&\ \ \ \ \ \ \ 
+ \delta_{i} C_{\max} \Big( \max_{i,c}\big\{ Q_{i}^{(c)}(t) \big\} + T\gamma_{\max} \Big) + T\Phi_{i} \Big]  
\bigg| \mathbf{X}(t) \bigg]  \nonumber \\
\leq& \bbbe \bigg[ 
T\Phi' - \epsilon \sum_{\tau=t}^{t+T-1} \sum_{i,c}Q_{i}^{(c)}(\tau) +TVh^*(\boldsymbol{\lambda} + \epsilon \mathbf{1})  \nonumber \\
&+ \sum_{ (i,j)\in\mathcal{E}} \Big[ \sum_{\tau=t}^{t+T} g\Big( C_{_{\max}} \max_{i,c}\big\{ Q_{i}^{(c)}(\tau) \big\} \Big)   \nonumber \\
&\ \ \ \ \ \ \ \ \ \ \ \ \ \ \ \ \ \ \ 
+  \delta_{ij} C_{_{\max}} \max_{i,c}\big\{ Q_{i}^{(c)}(t) \big\}  \Big] \nonumber \\
&+ 
\sum_{i \in \mathcal{V}} \Big[  \sum_{\tau=t}^{t+T} g\Big( C_{_{\max}} \max_{i,c}\big\{ Q_{i}^{(c)}(\tau) \big\} \Big) \nonumber \\
&\ \ \ \ \ \ \ \ \ \ \ \ \ \ \ \ \ \ \ 
+  \delta_{i} C_{_{\max}} \max_{i,c}\big\{ Q_{i}^{(c)}(t) \big\} \Big] 
\bigg| \mathbf{X}(t) \bigg] 
\end{align}
where 
$\Phi' = \Phi + \sum\limits_{(i,j)\in\mathcal{E}} (\Phi_{ij}+\delta_{ij}C_{\max}T\gamma_{\max} ) + \sum\limits_{i\in\mathcal{V}} (\Phi_{i} + \delta_{i}C_{\max}T\gamma_{\max})$.

Now select $T > \max \{ \frac{8|\mathcal{E}|\delta_{ij}C_{ij,\max}}{\epsilon}, \frac{8|\mathcal{V}|\delta_{i}C_{i,\max}}{\epsilon}\}$, and since $g$ is a sublinear function, there exists a constant $K_{T} < \infty$ such that $x > K_{T}$ implies $|\mathcal{E}|g(C_{\max} x) < \frac{\epsilon}{8}x$ for any $(i,j) \in \mathcal{E}$ and $|\mathcal{V}|g(C_{\max} x) < \frac{\epsilon}{8}x$ for any node $i\in\mathcal{V}$.

We thus have
\begin{align}
&\bbbe \left[ L(\mathbf{Q}(t+T)) - L(\mathbf{Q}(t)) + V\sum_{\tau=t}^{t+T} h(\tau) \bigg| \mathbf{X}(t) \right]  \nonumber \\
\leq& \bbbe \bigg[ 
T\Phi'' - \frac{\epsilon}{2} \sum_{\tau=t}^{t+T-1} \sum_{i,c} \bbbe[Q_{i}^{(c)}(\tau)] +TVh^*(\boldsymbol{\lambda} + \epsilon \mathbf{1}) 
\bigg| \mathbf{X}(t) \bigg]  \nonumber \\
\end{align}

Taking expectation on both sides and summing over time slots $t = 0, T, 2T, \dots, (K-1)T$, we have
\begin{align}
&\bbbe[ L(\mathbf{Q}(KT)) ] - \bbbe[ L(\mathbf{Q}(0)) ] + V\sum_{\tau=0}^{KT}\bbbe[h(\tau)]  \nonumber \\
\leq& KT\Phi'' - \frac{\epsilon}{2}\sum_{\tau=0}^{KT}\sum_{i,c}Q_{i}^{(c)}(\tau) + KTVh^*(\boldsymbol{\lambda}+\epsilon \mathbf{1})
\end{align}
Further divide both sides by $KT$, and rearranging the terms, we have
\begin{align}
&\frac{\epsilon}{2KT} \sum_{\tau=0}^{KT} \sum_{i,c} \bbbe[Q_{i}^{(c)}(\tau)] + \frac{V}{KT} \sum_{\tau=0}^{KT}\bbbe[h(\tau)]  \nonumber \\
\leq& \Phi'' + V h^*(\boldsymbol{\lambda}+\epsilon \mathbf{1}) + \frac{1}{2KT} \bbbe [\|\mathbf{Q}(0)\|^2]
\label{finaleq}
\end{align}

From \eqref{finaleq}, using~\cite[Prop. 6.1]{prob1_convergence}, Eqs. \eqref{eq8} and \eqref{eq9} in Theorem \ref{theorem_cost_delay} follow.
\end{proof}

\subsection{Proofs of Lemmas~\ref{lemma_change_bound}-\ref{lemma_resource_maximization}}

\begin{proof}{(of lemma~\ref{lemma_change_bound})}
First note that with assumption~\ref{assumption_arrivals}, for any queue $Q_{i}^{(c)}$, the queue length change over one time slot may be bounded as follows:
\begin{align*}
Q_{i}^{(c)}(\tau+1) - Q_{i}^{(c)}(\tau) \leq a_{\max} + C_{\max} (|\mathcal{V}^{-}(i)|+1)
\end{align*}
and 
\begin{align*}
Q_{i}^{(c)}(\tau+1) - Q_{i}^{(c)}(\tau) \geq -C_{\max} (|\mathcal{V}^{+}(i)|+1)
\end{align*}
Hence we have
\begin{align}
\Big| Q_{i}^{(c)}(\tau+1) - Q_{i}^{(c)}(\tau) \Big| \leq a_{\max} + C_{\max} (v_{\max}+1)
\end{align}
where $v_{\max} = \max\limits_{i} \{ \max\{|\mathcal{V}^+(i)|, |\mathcal{V}^-(i)|\}\}$.

By rearranging the terms and using the triangle inequality, we have
\begin{align}
&\Big| \frac{1}{\rho} \Big(Q_{i}^{(c_1)}(\tau+1) - \xi Q_{j}^{(c_2)}(\tau+1) \Big)   \nonumber \\
& \qquad \qquad \qquad \qquad \qquad \qquad  
- \frac{1}{\rho} \Big(Q_{i}^{(c_1)}(\tau) - \xi Q_{j}^{(c_2)}(\tau)\Big) \Big|  \nonumber \\
=&\Big| \frac{1}{\rho} \Big(Q_{i}^{(c_1)}(\tau+1) - Q_{j}^{(c_2)}(\tau+1) \Big) - \frac{\xi}{\rho} \Big(Q_{i}^{(c_1)}(\tau) - Q_{j}^{(c_2)}(\tau)\Big) \Big|  \nonumber \\
\leq& \frac{1}{\rho} \Big| Q_{i}^{(c_1)}(\tau+1) - Q_{i}^{(c_1)}(\tau)  \Big|  + \frac{\xi}{\rho} \Big|  Q_{j}^{(c_2)}(\tau+1) - Q_{j}^{(c_2)}(\tau) \Big|  \nonumber \\
\leq&  \frac{1}{\rho} (1+\xi) \Big( a_{\max} + C_{\max} (v_{\max}+1) \Big)  \nonumber\\
\leq& \frac{1}{\rho_{\min}} (1+\xi_{\max}) \big( a_{\max} + C_{\max} (v_{\max}+1) \big) 
\stackrel{\Delta}{=} \gamma_{\max}
\label{queue_length_change_bound}
\end{align}
which establishes the first part. 

Using (\ref{queue_length_change_bound}), we further prove the second part of the lemma. Let $c_{ij}^*(t)$ denote the commodity that maximizes the queue length differential at any time $t$, i.e. $c_{ij}^*(t) = \arg\max\limits_{c \in \mathcal{C}} \Big\{Q_{i}^{(c)}(\tau+1) - Q_{j}^{(c)}(\tau+1) \Big\}$, we then have
\begin{align}
&\max_{c \in \mathcal{C}} \Big\{Q_{i}^{(c)}(\tau+1) - Q_{j}^{(c)}(\tau+1) \Big\}  \nonumber \\
=& Q_{i}^{(c_{ij}^*(\tau+1))}(\tau+1) - Q_{j}^{(c_{ij}^*(\tau+1))}(\tau+1)  \nonumber \\
\leq&  \Big( Q_{i}^{(c_{ij}^*(\tau+1))}(\tau) - Q_{j}^{(c_{ij}^*(\tau+1))}(\tau) \Big) + \gamma_{\max}   \nonumber \\
\leq&  \max_{c \in \mathcal{C}} \Big\{Q_{i}^{(c)}(\tau) - Q_{j}^{(c)}(\tau) \Big\} + \gamma_{\max}
\end{align}
On the other hand, 
\begin{align}
&\max_{c \in \mathcal{C}} \Big\{Q_{i}^{(c)}(\tau+1) - Q_{j}^{(c)}(\tau+1) \Big\}  \nonumber \\
\geq& Q_{i}^{(c_{ij}^*(\tau))}(\tau+1) - Q_{j}^{(c_{ij}^*(\tau))}(\tau+1)  \nonumber \\
\geq&  \Big( Q_{i}^{(c_{ij}^*(\tau))}(\tau) - Q_{j}^{(c_{ij}^*(\tau))}(\tau) \Big) - \gamma_{\max}   \nonumber \\
=&  \max_{c \in \mathcal{C}} \Big\{Q_{i}^{(c)}(\tau) - Q_{j}^{(c)}(\tau) \Big\} - \gamma_{\max}
\end{align}
Combining the two bounds above, we have (\ref{eq:max_diff_tx_bound}). Similarly, let $c_{i}^*(t) = \arg\max\limits_{c \in \mathcal{C}} \Big\{ \frac{1}{\rho_{i}^{(c)}} \big[ Q_{i}^{(c)}(t) - \xi^{(c^+)} Q_{i}^{(c^+)}(t) -Ve_{i} \big]^+\Big\}$, we have 
\begin{align}
&\max_{c \in \mathcal{C}} \Big\{ \frac{1}{\rho_{i}^{(c)}} \big[ Q_{i}^{(c)}(\tau+1) - \xi^{(c^+)} Q_{i}^{(c^+)}(\tau+1) -Ve_{i} \big]^+\Big\}  \nonumber \\
=& \frac{1}{\rho_{i}^{(c_{i}^*(\tau+1))}} \big[ Q_{i}^{(c_{i}^*(\tau+1))}(\tau+1) - Q_{j}^{(c_{i}^*(\tau+1))}(\tau+1) -Ve_{i} \big]^+ \nonumber \\
\leq&  \frac{1}{\rho_{i}^{(c_{i}^*(\tau+1))}} \big[ Q_{i}^{(c_{i}^*(\tau+1))}(\tau) - Q_{j}^{(c_{i}^*(\tau+1))}(\tau) -Ve_{i} \big]^+ + \gamma_{\max}   \nonumber \\
\leq&  \max_{c \in \mathcal{C}} \Big\{ \frac{1}{\rho_{i}^{(c)}} \big[ Q_{i}^{(c)}(\tau) - \xi^{(c^+)} Q_{i}^{(c^+)}(\tau) -Ve_{i} \big]^+\Big\} + \gamma_{\max}
\end{align}
and
\begin{align}
&\max_{c \in \mathcal{C}} \Big\{ \frac{1}{\rho_{i}^{(c)}} \big[ Q_{i}^{(c)}(\tau+1) - \xi^{(c^+)} Q_{i}^{(c^+)}(\tau+1) -Ve_{i} \big]^+\Big\}  \nonumber \\
\geq& \frac{1}{\rho_{i}^{(c_{i}^*(\tau))}} \big[ Q_{i}^{(c_{i}^*(\tau))}(\tau+1) - Q_{j}^{(c_{i}^*(\tau))}(\tau+1) - Ve_{i} \big]^+ \nonumber \\
\geq&  \frac{1}{\rho_{i}^{(c_{i}^*(\tau))}} \big[ Q_{i}^{(c_{i}^*(\tau))}(\tau) - Q_{j}^{(c_{i}^*(\tau))}(\tau) -Ve_{i} \big]^+ - \gamma_{\max}   \nonumber \\
=&  \max_{c \in \mathcal{C}} \Big\{ \frac{1}{\rho_{i}^{(c)}} \big[ Q_{i}^{(c)}(\tau) - \xi^{(c^+)} Q_{i}^{(c^+)}(\tau) -Ve_{i} \big]^+\Big\} - \gamma_{\max}
\end{align}
We then have (\ref{eq:max_diff_px_bound}) from the two bounds above.

\end{proof}

\begin{proof}{(of lemma~\ref{lemma_resource_maximization})}

\begin{enumerate}
\item [(a)] Without loss of generality, assume that $y \geq x$. Let $k_y$ be the maximizer in the definition for $F(y)$, i.e. $F(y) = C_{ij}(k_y)[y-Ve_{ij}]^+ -Vw_{ij}(k_y)$.

Since $F(x) \geq C_{ij}(k_y)[x-Ve_{ij}]^+ -Vw_{ij}(k_y)$ by definition, we then have 
\begin{align}
F(y) - F(x) 
\leq F(y) - C_{ij}(k_y)[x-Ve_{ij}]^+ -Vw_{ij}(k_y)  \nonumber \\
\leq C_{ij}(k_y) (y-x)  \leq C_{ij,\max}(y-x)
\end{align}
The result then follows by the fact that $F(x)$ is strictly increasing.

\item[(b)]
If $\max\limits_{c\in\mathcal{C}} \big\{ Q_{i}^{(c)}(\tau)-Q_{j}^{(c)}(\tau) \} > V \big( \min\limits_{k>0}\frac{w_{ij}(k)}{C_{ij}(k)} + e_{ij} \big)$, let $k' > 0$ be the minimizer of $\min\limits_{k>0}\frac{w_{ij}(k)}{C_{ij}(k)}$. We then have
\begin{align}
&C_{ij}(k') \big[ \max\limits_{c\in\mathcal{C}} \big\{ Q_{i}^{(c)}(\tau)-Q_{j}^{(c)}(\tau) \} - Ve_{ij} \big]^+ - Vw_{ij}(k')  \nonumber \\
>& C_{ij}(k')  \Big( V \min_{k>0}\frac{w_{ij}(k)}{C_{ij}(k)} - V \frac{W_{ij}(k')}{C_{ij}(k')}  \Big) = 0
\end{align}
hence the weight maximizing resource allocation is nonzero.

On the other hand, if $\max\limits_{c\in\mathcal{C}} \big\{ Q_{i}^{(c)}(\tau)-Q_{j}^{(c)}(\tau) \} \leq V \big( \min\limits_{k}\frac{w_{ij}(k)}{C_{ij}(k)} + e_{ij} \big)$, then for any $k' > 0$, we have
\begin{align}
&C_{ij}(k') \big[ \max\limits_{c\in\mathcal{C}} \big\{ Q_{i}^{(c)}(\tau)-Q_{j}^{(c)}(\tau) \} - Ve_{ij} \big]^+ - Vw_{ij}(k')  \nonumber \\
\leq& C_{ij}(k')  \Big( V \min_{k>0}\frac{w_{ij}(k)}{C_{ij}(k)} - V \frac{W_{ij}(k')}{C_{ij}(k')}  \Big) \leq 0
\end{align}
Hence the resource allocation that maximizes the weight $\max\limits_{k} \big\{C_{ij}(k) \big[ \max\limits_{c\in\mathcal{C}} \big\{ Q_{i}^{(c)}(\tau)-Q_{j}^{(c)}(\tau) \} - Ve_{ij} \big]^+ - Vw_{ij}(k) \big\}$ is $k=0$.

\end{enumerate}

\end{proof}

\subsection{Proof of Theorem~\ref{theorem_capacity_region}}
\label{appendix_proof_capacity_region}


According to~\cite[Theorem 1]{dcnc_info16}, the capacity region with zero reconfiguration delay and cost, $\boldsymbol{\Lambda}$, consists of arrival rates $\lambda_{i}^{(c)}$ for which there exist multi-commodity flow variables $f_{ij}^{(c)}$, $f_{i}^{(c)}$ and probability values $\alpha_{i,k}, \alpha_{ij,k}, \beta_{i,k}^{(c)}, \beta_{ij,k}^{(c)}$ that satisfy:
\begin{align}
&\sum_{j \in \mathcal{V}^{-}(i)} f_{ji}^{(c)} + \xi^{(c)} f_{i}^{(c^-)} + \lambda_{i}^{(c)} \leq \sum_{j\in\mathcal{V}^{+}(i)} f_{ij}^{(c)} + f_{i}^{(c)}, \ \ \forall i, c   \label{eq:capacity_first}\\
&f_{i}^{(c)} \leq \frac{1}{\rho_{i}^{(c)}} \sum_{k \in \mathcal{K}_i} \alpha_{i,k} \beta_{i,k}^{(c)} C_{i}(k), \ \ \forall i, c  \label{eq:flow_variable1}\\
&f_{ij}^{(c)} \leq \frac{1}{\rho_{ij}^{(c)}} \sum_{k \in \mathcal{K}_ij} \alpha_{ij,k} \beta_{ij,k}^{(c)} C_{ij}(k), \ \ \forall (i,j), c \label{eq:flow_variable2}\\
&f_{i}^{(c)} \geq 0, \ \forall i, c, \ \ f_{ij}^{(c)} \geq 0, \ \ \forall (i,j), c \\
&\sum_{k \in K_{i}} \alpha_{i,k} \leq 1, \ \forall i, \ \ \sum_{k \in K_{ij}} \alpha_{ij,k} \leq 1, \ \ \forall (i,j) \\
&\sum_{c \in \mathcal{C}} \beta_{i}^{(c)} \leq 1, \ \ \forall i,  \ \ \sum_{c \in \mathcal{C}} \beta_{ij}^{(c)} \leq 1, \ \ \forall (i,j)  \label{eq:capacity_last}
\end{align}

Furthermore, the minimum average cloud network cost required for network stability under zero reconfiguration delay and cost is given by
\begin{align}
h^*(\boldsymbol{\lambda})
=& \min \sum_{i} \sum_{k \in \mathcal{K}_{i}} \alpha_{i,k} \Big( w_{i}(k) + e_i C_{i}(k) \sum_{c} \frac{\beta_{i,k}^{(c)}}{\rho_{i}^{(c)}} \Big)  \nonumber \\
&+ \sum_{(i,j)}\sum_{k \in \mathcal{K}_{ij}} \Big( w_{ij}(k) + e_{ij} C_{ij}(k) \sum_{c} \beta_{ij,k}^{(c)} \Big)
\end{align}
where the minimum is over all $f_{ij}^{(c)}$, $f_{i}^{(c)}$, $\alpha_{i,k}$, $\alpha_{ij,k}$, $\beta_{i,k}^{(c)}$, $\beta_{ij,k}^{(c)}$ that satisfy (\ref{eq:capacity_first})-(\ref{eq:capacity_last}).

The necessity of the Theorem (i.e. any rate $\boldsymbol{\lambda} \notin \boldsymbol{\Lambda}$ could not be stabilized by any policy) follows the same approach in the proof of~\cite[Theorem 1]{dcnc_info16}. Therefore, we have $\boldsymbol{\Lambda}_{\Delta} \subset \boldsymbol{\Lambda}$ and $h_{\Delta}^*(\boldsymbol{\lambda}) \geq h^*(\boldsymbol{\lambda})$ for any reconfiguration delay/cost structure $\Delta$. In the following, for each $\boldsymbol{\lambda} \in \boldsymbol{\Lambda}$ under any reconfiguration delay/cost structure $\Delta$, we construct policies that stabilize $\boldsymbol{\lambda}$ with average cloud network cost arbitrarily close to $h^*(\boldsymbol{\lambda})$.

We first establish that for each $\boldsymbol{\lambda} = \{\lambda_{i}^{(c)}\}$ that is in the interior of $\boldsymbol{\Lambda}$, i.e. $\boldsymbol{\lambda} \in \boldsymbol{\Lambda}^{o}$, there exists a policy that stabilizes the cloud network under the exogenous arrival rate $\boldsymbol{\lambda}$ and under any finite reconfiguration delay and cost, $\Delta = \Big\{ \{\delta_{i}\}_{i\in\mathcal{V}}, \{\delta_{ij}\}_{(i,j)\in\mathcal{E}}, \{\eta_{i}\}_{i\in\mathcal{V}}, \{\eta_{ij}\}_{(i,j)\in\mathcal{E}} \Big\}$.
Notice that since reconfiguration cost does not affect the queue dynamics as in (\ref{queue_dynamics}), hence we may ignore the reconfiguration cost when considering cloud network stability. We also denote the maximum reconfiguration delay as $\delta_{\max} = \max\{ \max\limits_{i \in \mathcal{V}} \delta_{i}, \max\limits_{(i,j)\in\mathcal{E}} \delta_{ij} \}$.

Since $\boldsymbol{\lambda} \in \boldsymbol{\Lambda}^{o}$, there exists $\epsilon, \epsilon' > 0$ such that $(1+\epsilon') \boldsymbol{\lambda} + \epsilon \mathbf{1} \in \boldsymbol{\Lambda}^{o}$. Then substituting $\boldsymbol{\lambda}$ as $(1+\epsilon') \boldsymbol{\lambda} + \epsilon \mathbf{1}$ in (\ref{eq:capacity_first}), there exist variables $f_{ij}^{(c)}$, $f_{i}^{(c)}$, $\alpha_{i,k}$, $\alpha_{ij,k}$, $\beta_{i,k}^{(c)}$, $\beta_{ij,k}^{(c)}$ that satisfy (\ref{eq:capacity_first})-(\ref{eq:capacity_last}). 



At each time instance, we may concatenate the flow allocation variables $\{\mu_{i}^{(c)}(t)\}_{i,c}$, $\{\mu_{ij}^{(c)}(t)\}_{(i,j),c}$, and get a $L = (|\mathcal{V}|+|\mathcal{E}|)|\mathcal{C}|$ dimensional vector $\boldsymbol{\mu}(t) = [\mu_{i}^{(c)}(t); \mu_{ij}^{(c)}(t)] \in \bbbr^L$.
Let $\mathcal{M}$ be the set of all feasible $L$-dimensional flow allocations, i.e. $\boldsymbol{\mu}(t) \in \mathcal{M} \subset \bbbr^{L}$. 

Now concatenating flow variables $\{f_{i}^{(c)}\}_{i,c}$ and $\{f_{ij}^{(c)}\}_{(i,j),c}$ as $\mathbf{f} = [f_{i}^{(c)}; f_{ij}^{(c)}] \in \bbbr^L$, it is straightforward from (\ref{eq:flow_variable1}) and (\ref{eq:flow_variable2}) that $\mathbf{f}$ is in the convex hull of $\mathcal{M}$. Hence according to Caratheodory's Theorem, $\mathbf{f}$ could be decomposed as a convex combination of $L+1$ vectors $\{\boldsymbol{\mu}_{l}\}_{l=1}^{L+1}$ in $\mathcal{M}$. 

\begin{align}
\mathbf{f} = \sum_{l=1}^{L+1} \gamma_{l} \boldsymbol{\mu}_{l}
\label{load_decomposition}
\end{align}

With the decomposition given in (\ref{load_decomposition}), we may construct a periodic flow allocation schedule of period $T$ such that each vector $\boldsymbol{\mu}_{l}$ in the $L+1$ vectors $\{\boldsymbol{\mu}_{l}\}_{l=1}^{L+1}$ is actively scheduled for $\gamma_{l} (T - (L+1)\delta_{\max})$ time slots within one period. In particular, each $\boldsymbol{\mu}_{l}$ is scheduled for consecutive $\gamma_{l} (T - (L+1)\delta_{\max}) + \delta_{\max}$ time slots, with the first $\delta_{\max}$ being reserved for reconfiguration. We then have $\sum_{t=kT}^{(k+1)T-1} \mu_{ij}^{(c)}(t) \mathds{1}_{\{r_{ij}(t)=0\}} = (T - (L+1)\delta_{\max}) f_{ij}^{(c)}$ and $\sum_{t=kT}^{(k+1)T-1} \mu_{i}^{(c)}(t) \mathds{1}_{\{r_{i}(t)=0\}} = (T - (L+1)\delta_{\max}) f_{i}^{(c)}$. Now consider the Lyapunov function $L(\mathbf{Q}) = \sum_{i,c}Q_{i}^{(c)}$, and we have that for each node $i$ and each commodity $c$ that:



\begin{align}
&\bbbe \bigg[ Q_{i}^{(c)}(t+T) - Q_{i}^{(c)}(t) \bigg| \mathbf{X}(t) \bigg]  \nonumber\\
\leq& T\lambda_{i}^{(c)} + \sum_{\tau=t}^{t+T-1} \bbbe \bigg[ -\sum_{j\in\mathcal{V}^{+}(i)} \mu_{ij}^{(c)}(\tau) \mathds{1}_{\{r_{ij}(t)=0\}} - \mu_{i}^{(c)}(\tau)\mathds{1}_{\{r_{i}(\tau)=0\}}  \nonumber \\
&+ \sum_{j\in\mathcal{V}^{-}(i)} \mu_{ji}^{(c)}\mathds{1}_{\{r_{ji}(\tau)=0\}} + \xi_{i}^{(c)}\mu_{i}^{(c^-)}(\tau)\mathds{1}_{\{r_{i}(\tau)=0\}} \bigg| \mathbf{X}(t) \bigg]  \nonumber \\
=& T\lambda_{i}^{(c)} + \Big( T - (L+1)\delta_{\max} \Big) \Big[ -\sum_{j\in\mathcal{V}^{+}(i)} f_{ij}^{(c)} - f_{i}^{(c)}(\tau) \nonumber \\
&+ \sum_{j\in\mathcal{V}^{-}(i)} f_{ji}^{(c)} + \xi_{i}^{(c)}f_{i}^{(c^-)}(\tau) \Big]  \nonumber \\
\leq& T\lambda_{i}^{(c)} - \Big( T - (L+1)\delta_{\max} \Big) \Big[ (1+\epsilon')\lambda_{i}^{(c)} - \epsilon \Big]  
\end{align}
Then for any $T$ that satisfies $( 1 - \frac{(L+1)\delta_{\max}}{T} ) (1+\epsilon') > 1$, or in other words, any $T > (1 + \frac{1}{\epsilon'})(L+1)\delta_{\max}$, we have

\begin{align}
&\bbbe \bigg[ Q_{i}^{(c)}(t+T) - Q_{i}^{(c)}(t) \bigg| \mathbf{X}(t) \bigg] 
\leq - \Big( T - (L+1)\delta_{\max} \Big) \epsilon
\label{eq:periodic_neg_drift}
\end{align}
The rate stability then follows from Foster-Lyapunov Theorem.

For the average cloud network cost, with the solution $f_{ij}^{(c)}$, $f_{i}^{(c)}$, $\alpha_{i,k}$, $\alpha_{ij,k}$, $\beta_{i,k}^{(c)}$, $\beta_{ij,k}^{(c)}$ of (\ref{eq:capacity_first})-(\ref{eq:capacity_last}), we may similarly construct a periodic flow allocation schedule that achieves arbitrarily close to the minimal average cloud network cost $h^*(\boldsymbol{\lambda})$. In particular, following the same construction, a periodic flow schedule of period $T$ achieves average cost of $h^*(\boldsymbol{\lambda}) + \frac{L+1}{T}(\sum_{i\in\mathcal{V}} \eta_{i} + \sum_{(i,j)\in\mathcal{E}} \eta_{ij})$. Since we may select $T$ arbitrarily large without affecting the stability (as long as $T > (1 + \frac{1}{\epsilon'})(L+1)\delta_{\max}$), as shown previously in (\ref{eq:periodic_neg_drift}), we have that $h_{\Delta}^*(\boldsymbol{\lambda}) = h^*(\boldsymbol{\lambda})$

\end{document}